\documentclass[11pt]{livrev}
\usepackage{epsf}

\newcommand{\mathbold}[1]{\mbox{\boldmath $#1$}}

\newlength{\zerolength}
\newcommand{\blankzero}{\hspace{\zerolength}}

\newlength{\dotlength}
\newcommand{\blankdot}{\hspace{\dotlength}}

\newlength{\gtlength}
\newcommand{\blankgt}{\hspace{\gtlength}}

\begin{document}

\title{Relativistic Binaries in Globular Clusters}

\author{Matthew  J.\ Benacquista \\
		Montana State University-Billings \\
		1500 N.\ 30th \\
		Billings, Montana 59101 \\
		e-mail: benacquista@msubillings.edu \\
		http://www.msubillings.edu/ScienceFaculty/Benacquista}

\date{}

\maketitle


\begin{abstract}
  The galactic population of globular clusters are old, dense star
  systems, with a typical cluster containing $10^4 - 10^7$ stars. As
  an old population of stars, globular clusters contain many collapsed
  and degenerate objects. As a dense population of stars, globular
  clusters are the scene of many interesting close dynamical
  interactions between stars. These dynamical interactions can alter
  the evolution of individual stars and can produce tight binary
  systems containing one or two compact objects. In this review, we
  discuss the theoretical models of globular cluster evolution and
  binary evolution, techniques for simulating this evolution which
  lead to relativistic binaries, and current and possible future
  observational evidence for this population. Globular cluster
  evolution will focus on the properties that boost the production of
  hard binary systems and on the tidal interactions of the galaxy with
  the cluster, which tend to alter the structure of the globular
  cluster with time. The interaction of the components of hard binary
  systems alters the evolution of both bodies and can lead to exotic
  objects. Direct $N$-body integrations and Fokker--Planck simulations
  of the evolution of globular clusters that incorporate tidal
  interactions and lead to predictions of relativistic binary
  populations are also discussed. We discuss the current observational
  evidence for cataclysmic variables, millisecond pulsars, and
  low-mass X-ray binaries as well as possible future detection of
  relativistic binaries with gravitational radiation.
\end{abstract}


\newpage


\section{Introduction}
\label{section:introduction}

Relativistic binaries containing white dwarfs, neutron stars, and black holes in compact orbits are over-represented in globular clusters compared with their population in the galactic field. Observations of this population reveal a host of exotic objects such as ultracompact cataclysmic variables, non-flickering X-ray and UV sources, low-mass X-ray binaries, and millisecond pulsars. These objects and their dark counterparts in the population of relativistic binaries are also likely to be observable sources of gravitational radiation for low-frequency gravitational wave detectors such as the planned space-borne interferometer, LISA. In the field, a relativistic binary is a product of the interplay between stellar evolution and the gravitational interaction of a tight binary. In globular clusters, the population of tight binaries is also a product of the dynamical evolution of an $N$-body gravitational system. Thus, relativistic binaries result from a combination of several of the more interesting processes in astrophysics. In keeping with the focus of this review article on relativistic binaries in globular clusters, we shall only touch on the aspects of globular clusters, observations, binary evolution, and $N$-body dynamics as they relate to populations of this specific class of binaries in globular clusters.

We begin by looking at the physical structure and general history of
the galactic globular cluster system that leads to the concentration
of evolved stars, stellar remnants, and binary systems in the cores of
the clusters. Current observations of globular clusters
that have revealed numerous populations of relativistic
binaries and their tracers are also presented. We also look at the prospects for future
observations in this rapidly changing area. Many of these relativistic
binaries are the product of stellar evolution in compact binaries. We
will look at how mass transfer from one star in the presence of a
nearby companion can dramatically alter the evolution of both stars in
the process of binary evolution. The enhanced production of
relativistic binaries in globular clusters results from dynamical
processes that drive binaries toward tighter orbits and that
preferentially exchange more massive and degenerate objects into
binary systems. Numerical simulations of globular cluster evolution,
which can be used to predict the rate at which relativistic binaries
are formed, are discussed. These models are compared with the
observable members of the population of relativistic
binaries. Finally, we conclude with a brief discussion of the
prospects for observing these systems in gravitational radiation.

Readers interested in further studies of the structure and evolution
of globular clusters are invited to look at Binney and Tremaine~\cite{binney87}, Spitzer~\cite{spitzer87}, and Volumes I and
II of Padmanabhan's {\it Theoretical Astrophysics}~\cite{padmanabhan00, padmanabhan01} for a good introduction to the physical processes involved. Review articles of Meylan \& Heggie~\cite{meylan97} and Meylan~\cite{meylan99} also provide a comprehensive look at the internal dynamics of globular clusters. Although our focus is solely on the Galactic globular cluster system, the physics of globular cluster systems associated with other galaxies is well covered in the review article by Harris~\cite{harris91} as well as his lecture notes from the Saas-Fee course on star clusters~\cite{harris01}. Carney has a thorough introduction to evolution of stars in globular clusters~\cite{carney01}. An observational perspective on the role of binaries in globular clusters is presented in an excellent review by Bailyn~\cite{bailyn95}, while a good introduction to the details of observing binary systems in general can be found in {\it An Introduction to Close Binary Stars}~\cite{hilditch01}. Although slightly out of date, the review of binaries in globular clusters by Hut {\it et al.}~\cite{hut92a} is an excellent introduction to the interaction between globular cluster dynamics and binary evolution, as is a short article on globular cluster binaries by McMillan, Pryor, and Phinney~\cite{mcmillan98}. Rappaport {\it  et al.}~\cite{rappaport01} and Rasio {\it et al.}~\cite{rasio01} have written reviews of numerical simulations of binary populations in globular clusters. An excellent introduction to the astrophysics and numerical techniques relevant to globular cluster dynamics can be found in the book by Heggie and Hut~\cite{heggie03}.
\newpage


\section{Globular Clusters}
\label{section:globular_clusters}

\begin{figure}[htb]
  \def\epsfsize#1#2{0.7#1}
  \centerline{\epsfbox{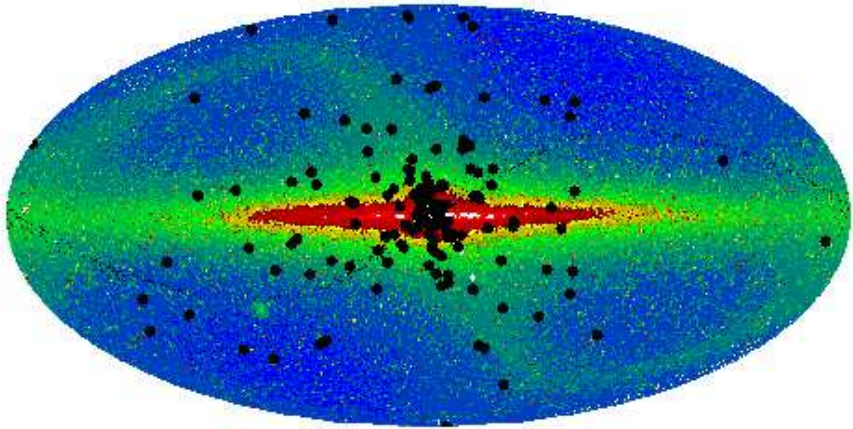}}
  \caption{\it Globular cluster distribution about the galaxy. Positions
    are from Harris\protect~\protect\cite{harris96} and are plotted as
    black circles on top of the COBE FIRAS 2.2 micron map of the
    Galaxy using a Mollweide projection. Figure taken from Brian
    Chaboyer's website\protect~\protect\cite{chaboyerweb}.}
\label{GC_coordinates}
\end{figure}

Globular clusters comprising $10^4$ to $10^7$ stars were formed early
in the history of the Milky Way, and are scattered throughout the
halo. The age of the clusters is about 13~Gyr, with an age spread of
less than 5~Gyr~\cite{carretta00}. According to the frequently updated
catalog of globular cluster properties maintained on the web by
Harris~\cite{harris96}, the globular cluster system numbers roughly
150 clusters. Although the list is fairly complete, new globular
clusters have been recently discovered at very low galactic
latitudes~\cite{hurt00, kobulnicky05}, and there is the prospect for a few more clusters to be hidden behind the bulge or out in the far reaches of the galaxy. The distribution of globular clusters in galacto-centric coordinates is shown in Figure~\ref{GC_coordinates}.

\subsection{Globular cluster stars}
\label{subsection:globular_cluster_stars}

\begin{figure}[htb]
  \def\epsfsize#1#2{1.0#2}
  \centerline{\epsfbox{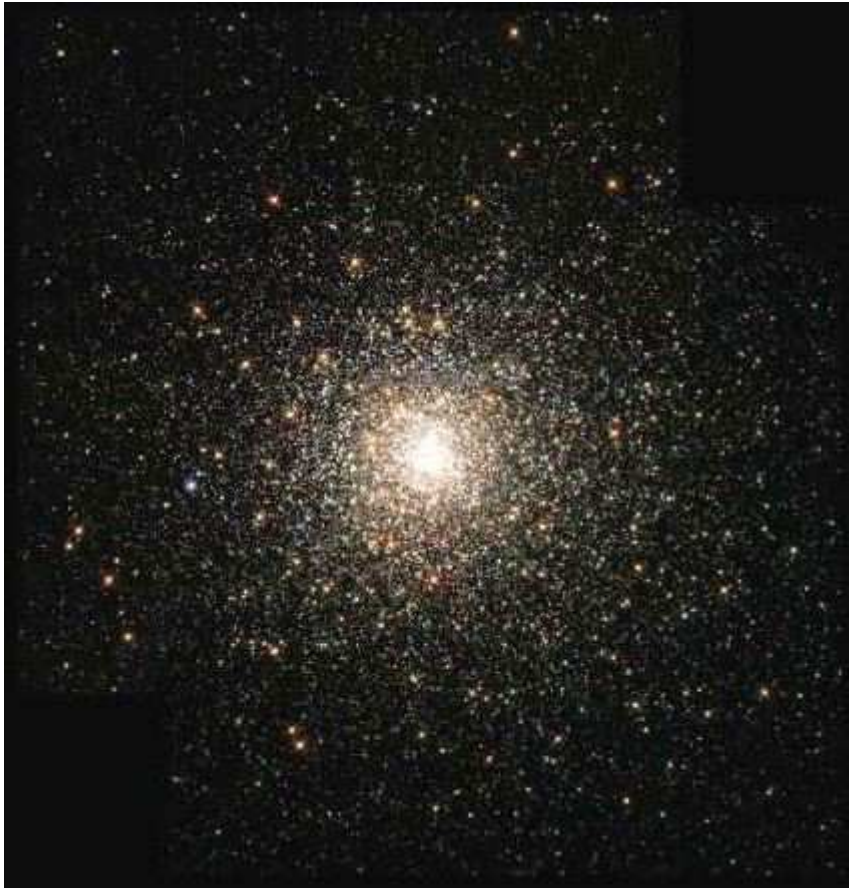}}
  \caption{\it Hubble Space Telescope photograph of the dense globular
    cluster M80 (NGC 6093).}
  \label{M80}
\end{figure}

\begin{figure}[htb]
  \def\epsfsize#1#2{0.6#1}
  \centerline{\epsfbox{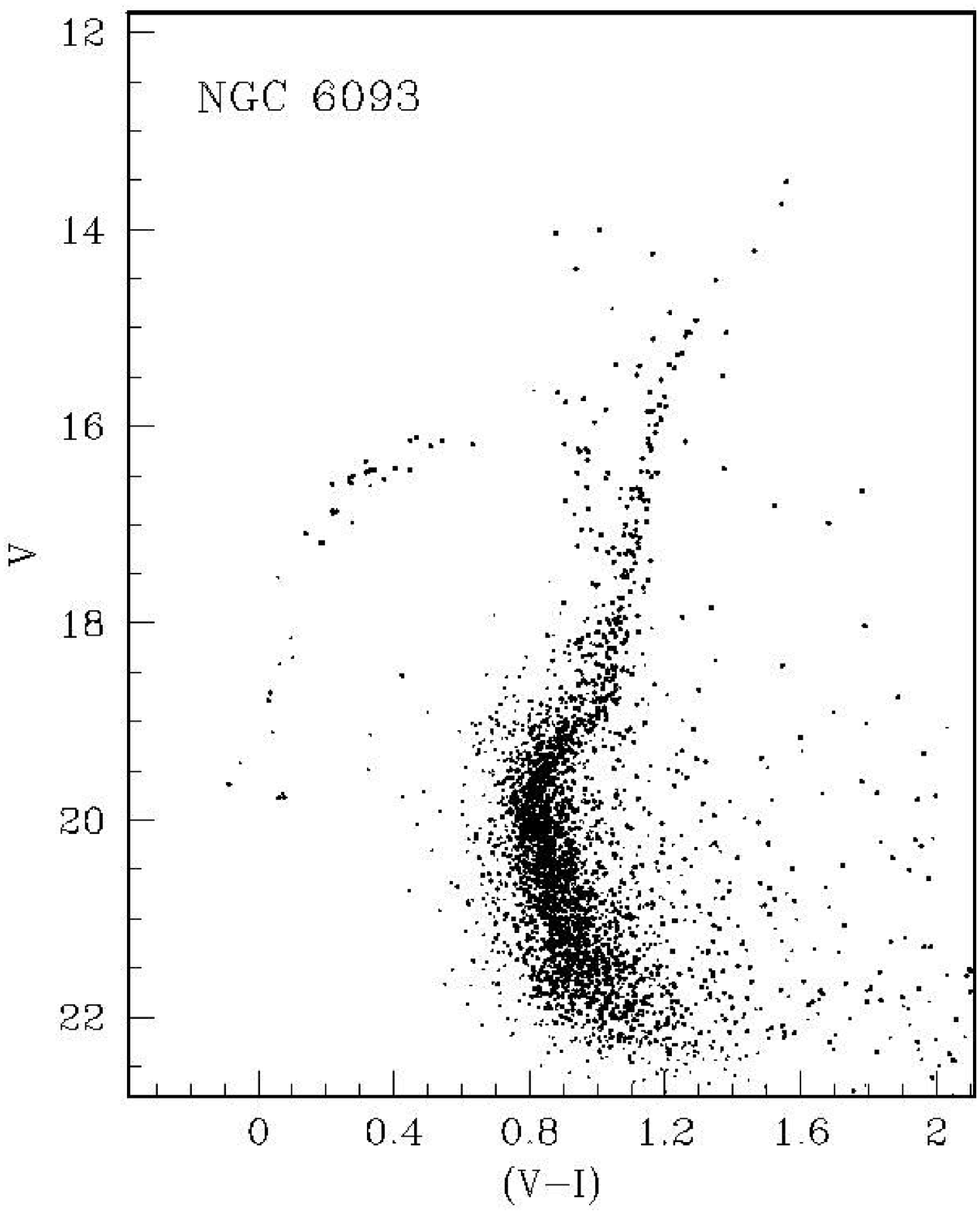}}
  \caption{\it Color--magnitude diagram for M80. Figure taken from the catalog of 52 globular clusters~\cite{rosenberg00}. The entire catalog is available at the Padova Globular Cluster Group website (http://dipastro.pd.astro.it/globulars/).}
  \label{M80_CMD}
\end{figure}

Because the clusters are of great age, most of the stars above about
$0.8 M_{\odot}$ have already evolved off the main sequence. Thus, a
large number of red giants are readily visible in most pictures of
globular clusters (see Figure~\ref{M80}).
When viewing the color--magnitude diagram (CMD) for a globular cluster,
one can clearly see the red giant branch lifting up away from the main
sequence. The horizontal branch of evolved stars is also seen in the
CMD for M80 shown in Figure~\ref{M80_CMD}. 

The clearly visible turn-off point in the CMD for globular
clusters is evidence for the roughly coeval nature of the stars in the
cluster. During the early stages of the evolution of a globular
cluster, most of the gas and dust has been swept away. Subsequent
replenishment of the intercluster gas by stellar winds from evolved
stars is removed during periodic passages of the cluster through the
plane of the galaxy. The remaining gas is generally too hot for any
star formation to take place~\cite{freire01a}. Thus globular clusters
are made up of old, population II stars.


\subsection{Globular cluster structure}
\label{subsection:globular_cluster_structure}

The overall structure of a globular cluster can be described in terms
of a roughly spherical $N$-body system with central densities in the range $10^{-1}$ to
$10^6 M_{\odot}/{\rm pc}^3$, and an average of
$10^4 M_{\odot}/{\rm pc}^3$. The important characteristic radii of a
globular cluster are the core radius $r_{\rm c}$, the half-light
radius $r_{\rm h}$, and the tidal radius $r_{\rm t}$. The core radius
is defined to be the radius at which the surface brightness has
dropped to half the central value. The half-light radius is the radius
that contains half of the light of the cluster and the tidal radius is
defined as the radius beyond which the external gravitational field of
the galaxy dominates the dynamics. Theorists define $r_{\rm h}$ to be
the radius containing half the mass of the cluster. The half-mass
radius is a three-dimensional theoretical construct, while the half-light radius
is a two-dimensional observational construct. The tidal radius is always determined by some
theoretical model. Typical values of these radii are 1.5~pc, 10~pc,
and 50~pc, respectively~\cite{binney87, padmanabhan01}.

There are also important characteristic time scales that govern the
dynamics of globular clusters. These are the crossing time
$t_{\rm cross}$, the relaxation time $t_{\rm relax}$, and the
evaporation time $t_{\rm evap}$. The crossing time is the typical time required for a star in the cluster to travel the characteristic size
$R$ of the cluster (typically taken to be the half-mass
radius). Thus, $t_{\rm cross} \sim R/v$, where $v$ is a typical
velocity ($\sim$\,10~km/s). The relaxation time is the typical
time for gravitational interactions with other stars in the cluster to
remove the history of a star's original velocity. This amounts to the
time required for gravitational encounters to alter the star's velocity
by an amount comparable to its original velocity. Since the relaxation
time is related to the number and strength of the gravitational
encounters of a typical cluster star, it is related to the number of
stars in the cluster and the average energy of the stars in the
cluster. Thus, it can be shown that the mean relaxation time for a
cluster is~\cite{binney87, padmanabhan00}
\begin{equation}
  t_{\rm relax} \simeq \frac{0.1 N}{\ln{N}} t_{\rm cross}.
  \label{relaxation_time}
\end{equation}
For a globular cluster with $N = 10^5$, a characteristic size of $R \sim r_h \sim 10~{\rm pc}$, and a typical velocity of $v \sim 10~{\rm km/s}$, the crossing time and relaxation time are $t_{\rm cross} \sim 10^6~{\rm yr}$ and $t_{\rm relax} \sim 10^9~{\rm yr}$, although Binney and Tremaine use $t_{\rm cross} \sim 10^5~{\rm yr}$ and consequently $t_{\rm relax} ~ 10^8~{\rm yr}$~\cite{binney87}. In real globular clusters, the relaxation time varies throughout the cluster and the median value is closer to $10^9~{\rm yr}$~\cite{binney87} as found in  Figure~1.3 of Spitzer~\cite{spitzer87} and Padmanabhan~\cite{padmanabhan01}.

The evaporation time for a cluster is the time required for the
cluster to dissolve through the gradual loss of stars that gain
sufficient velocity through encounters to escape its gravitational
potential. In the absence of stellar evolution and tidal interactions
with the galaxy, the evaporation time can be estimated by assuming
that a fraction $\gamma$ of the stars in the cluster are evaporated
every relaxation time. Thus, the rate of loss is
$dN/dt = -\gamma N/t_{\rm relax} = -N/t_{\rm evap}$. The value of
$\gamma$ can be determined by noting that the escape speed
$v_{\rm e}$ at a point $\mathbold{x}$ is related to the gravitational
potential $\Phi(\mathbold{x})$ at that point by
$v_{\rm e}^2 = -2\Phi(\mathbold{x})$. Consequently, the mean-square
escape speed in a cluster with density $\rho(\mathbold{x})$ is
\begin{equation}
  \langle{v^2_{\rm e}}\rangle = \frac{\int{\rho(\mathbold{x})
  v^2_{\rm e} d^3x}}{\int{\rho(\mathbold{x}) d^3x}} = -2
  \frac{\int{\rho(\mathbold{x}) \Phi(\mathbold{x}) d^3x}}{M} =
  -\frac{4W}{M},
  \label{escape_speed_1}
\end{equation}
where $W$ is the total potential energy of the cluster and $M$ is its
total mass. If the system is virialized (as we would expect after a
relaxation time), then $-W = 2K = M\langle{v^2}\rangle$, where $K$ is
the total kinetic energy of the cluster, and
\begin{equation}
  \langle{v^2_{\rm e}}\rangle = 4\langle{v^2}\rangle.
  \label{escape_speed_2}
\end{equation}
Thus, stars with speeds above twice the RMS speed will
evaporate. Assuming a Maxwellian distribution of speeds, the fraction
of stars with
$v > 2 v_{\rm rms}$ is $\gamma = 7.38 \times 10^{-3}$. Therefore, the
evaporation time is
\begin{equation}
  t_{\rm evap} = \frac{t_{\rm relax}}{\gamma} = 136 \, t_{\rm relax}.
  \label{evaporation_time}
\end{equation}
Stellar evolution and tidal interactions tend to shorten the
evaporation time. See Gnedin and Ostriker~\cite{gnedin97} and
references therein for a thorough discussion of these effects. Using a typical $t_{\rm relax}$ for a
globular cluster, we see that $t_{\rm evap} \sim 10^{10} {\rm\ yr}$, which is
comparable to the observed age of globular clusters.

The characteristic time scales of globular clusters differ significantly from each other:
$t_{\rm cross} \ll t_{\rm relax} \ll t_{\rm evap}$.

When discussing
stellar interactions during a given epoch of globular cluster
evolution, it is possible to describe the background structure of the
globular cluster in terms of a static model. These models describe the
structure of the cluster in terms of a distribution function $f$ that
can be thought of as providing a probability of finding a star at a
particular location in phase-space. The static models are valid over
time scales which are shorter than the relaxation time so that
gravitational interactions do not have time to significantly alter the
distribution function. We can therefore assume
$\partial f/\partial t \sim 0$. The structure of the globular cluster
is then determined by the collisionless Boltzmann equation,
\begin{equation}
  \mathbold{v} \cdot \mathbold{\nabla} f - \mathbold{\nabla} \phi
  \cdot \frac{\partial f}{\partial \mathbold{v}} = 0,
  \label{collisionless_boltzmann}
\end{equation}
where the gravitational potential $\phi$ is found from $f$ with
\begin{equation}
  \nabla^2\phi = 4\pi\int f(\mathbold{x}, \mathbold{v}, m) \, d^3v \,
dm.
  \label{static_potential}
\end{equation}

The solutions to Equation~(\ref{collisionless_boltzmann}) are often
described in terms of the relative energy per unit mass
${\cal E} \equiv \Psi - v^2/2$ with the relative potential defined as
$\Psi \equiv -\phi + \phi_0$. The constant $\phi_0$ is chosen so that
there are no stars with relative energy less than 0 (i.e.\ $f > 0$ for
${\cal E} > 0$ and $f = 0$ for ${\cal E} < 0$). A simple class of
solutions to Equation~(\ref{collisionless_boltzmann}),
\begin{equation}
  f({\cal E}) = F {\cal E}^{7/2},
  \label{plummer}
\end{equation}
generate what are known as Plummer models. A convenient class of
models which admit anisotropy and a distribution in angular momenta
$L$ are known as King--Michie models. The King--Michie distribution
function is:
\begin{equation}
  f({\cal E}, L) = \rho_1(2\pi \sigma^2)^{-3/2}
  \exp \left( \frac{-L^2}{2r_{\rm a}^2\sigma^2} \right)
  \left[e^{{\cal E}/{\sigma^2}}-1\right],
  \qquad {\cal E} > 0,
  \label{king-michie}
\end{equation}
with $f = 0$ for ${\cal E} \leq 0$ and $\rho_1$ a constant. The
velocity dispersion is determined by $\sigma$ and the anisotropy
radius $r_{\rm a}$ is defined so that the velocity distribution
changes from nearly isotropic at the center to nearly radial at
$r_{\rm a}$. The King--Michie distribution can be generalized to multi-mass systems, and although not dynamically correct, they can be used for mass estimates. A good description of the construction of a multi-mass King--Michie model can be found in the appendix of Miocchi~\cite{miocchi05}.


\subsection{Globular cluster evolution}
\label{subsection:globular_cluster_evolution}

An overview of the evolution of globular clusters can be found in Hut
{\it et al.}~\cite{hut92a}, Meylan \& Heggie~\cite{meylan97}, and
Meylan~\cite{meylan99}. We summarize here the aspects of globular
cluster evolution that are relevant to the formation and concentration
of relativistic binaries. The formation of globular clusters is not
well understood~\cite{geyer02} and the details of the initial mass
function (IMF) are an ongoing field of star cluster studies. The Kroupa mass function~\cite{kroupa01d} is the most common IMF currently used (see~\cite{kroupa01a} for a discussion of the local IMF). It has the form:
\begin{equation}
  dN \propto m^{-\alpha_i} dm.
  \label{imf}
\end{equation}
where
\begin{equation}
          \begin{array}{l@{\quad\quad,\quad}l}
\alpha_0 = +0.3\pm0.7   &0.01 \le m/M_\odot < 0.08, \\
\alpha_1 = +1.3\pm0.5   &0.08 \le m/M_\odot < 0.50, \\
\alpha_2 = +2.3\pm0.3   &0.50 \le m/M_\odot < 1.00, \\
\alpha_3 = +2.3\pm0.7   &1.00 \le m/M_\odot,\\
          \end{array}
\label{alphas}
\end{equation}
Some older work uses the Salpeter IMF which assumes a single value of $\alpha$ in Equation~(\ref{imf}) for all masses.
Once the stars form out of the initial molecular cloud the system is not virialized (i.e. it does not satisfy Equation~(\ref{collisionless_boltzmann}), and it
will undergo what is known as violent relaxation as the protocluster first begins to
collapse. During violent relaxation, the total energy of individual stars can change as the local gravitational potential changes. The process of violent relaxation is a collisionless process and it occurs rapidly over a timescale given by a few crossing times. During violent relaxation, the energy per mass of a given star changes in a way that is independent of the mass of the star~\cite{binney87}. Thus, the more massive stars will have more kinetic
energy. These stars will then lose their kinetic energy to the less
massive stars through stellar encounters, which leads towards
equipartition of energy. Through virialization, this tends to
concentrate the more massive stars in the center of the cluster---a process known as mass segregation. The
process of mass segregation for stars of mass $m_i$ occurs on a
timescale given by $t_i \sim t_{\rm relax} \langle{m}\rangle/m_i$.

The higher concentration of stars in the center of the cluster
increases the probability of an encounter, which, in turn, decreases
the relaxation time. Thus, the relaxation time given in
Equation~(\ref{relaxation_time}) is an average over the whole
cluster. The local relaxation time of the cluster is given in Meylan
\& Heggie~\cite{meylan97} and can be described by:
\begin{equation}
  t_{\rm r} = \frac{0.065{\langle{v^2}\rangle}^{3/2}}
  {\rho \langle{m}\rangle G^2 \ln{\Lambda}},
  \label{local_relaxation_time}
\end{equation}
where $\rho$ is the local mass density, $\langle{v^2}\rangle$ is the
mass-weighted mean square velocity of the stars, $\langle{m}\rangle$
is the mean stellar mass. The Coulomb logarithm, $\ln{\Lambda}$, is the logarithm of the ratio of the maximum to minimum expected impact parameters in the cluster. Typical values of $\Lambda$ range between $0.4 N$~\cite{meylan97,padmanabhan01} and $0.1 N$~\cite{giersz96}. Binney and Tremaine provide a range of values for $\ln{\Lambda}$ from 10.1 in the center of the cluster to 12 at $r_{h}$. Note that in the
central regions of the cluster, the value of $t_{\rm r}$ is much lower
than the average relaxation time. This means that in the core of the
cluster, where the more massive stars have concentrated, there are
more encounters between these stars.

The concentration of massive stars in the core of the cluster will
occur within a few relaxation times,
$t \sim t_{\rm relax} \sim 10^9 {\rm\ yr}$. This time is longer than
the lifetime of low metallicity stars with
$M \geq 2 M_{\odot}$~\cite{schaller92}. Consequently, these stars will
have evolved into carbon-oxygen (CO) and oxygen-neon (ONe) white
dwarfs, neutron stars, and black holes. After a few more
relaxation times, the average mass of a star in the globular cluster
will be around $0.5 M_{\odot}$ and these degenerate objects will once
again be the more massive objects in the cluster, despite having lost
most of their mass during their evolution. Thus, the population in the
core of the cluster will be enhanced in degenerate objects. Any
binaries in the cluster that have a gravitational binding energy
significantly greater than the average kinetic energy of a cluster
star will act effectively as single objects with masses equal to their
total mass. These objects, too, will segregate to the central regions
of the globular cluster~\cite{vesperini97}. The core will then be
overabundant in binaries and degenerate objects.

\begin{figure}[htbp]
  \def\epsfsize#1#2{0.6#1}
  \centerline{\epsfbox{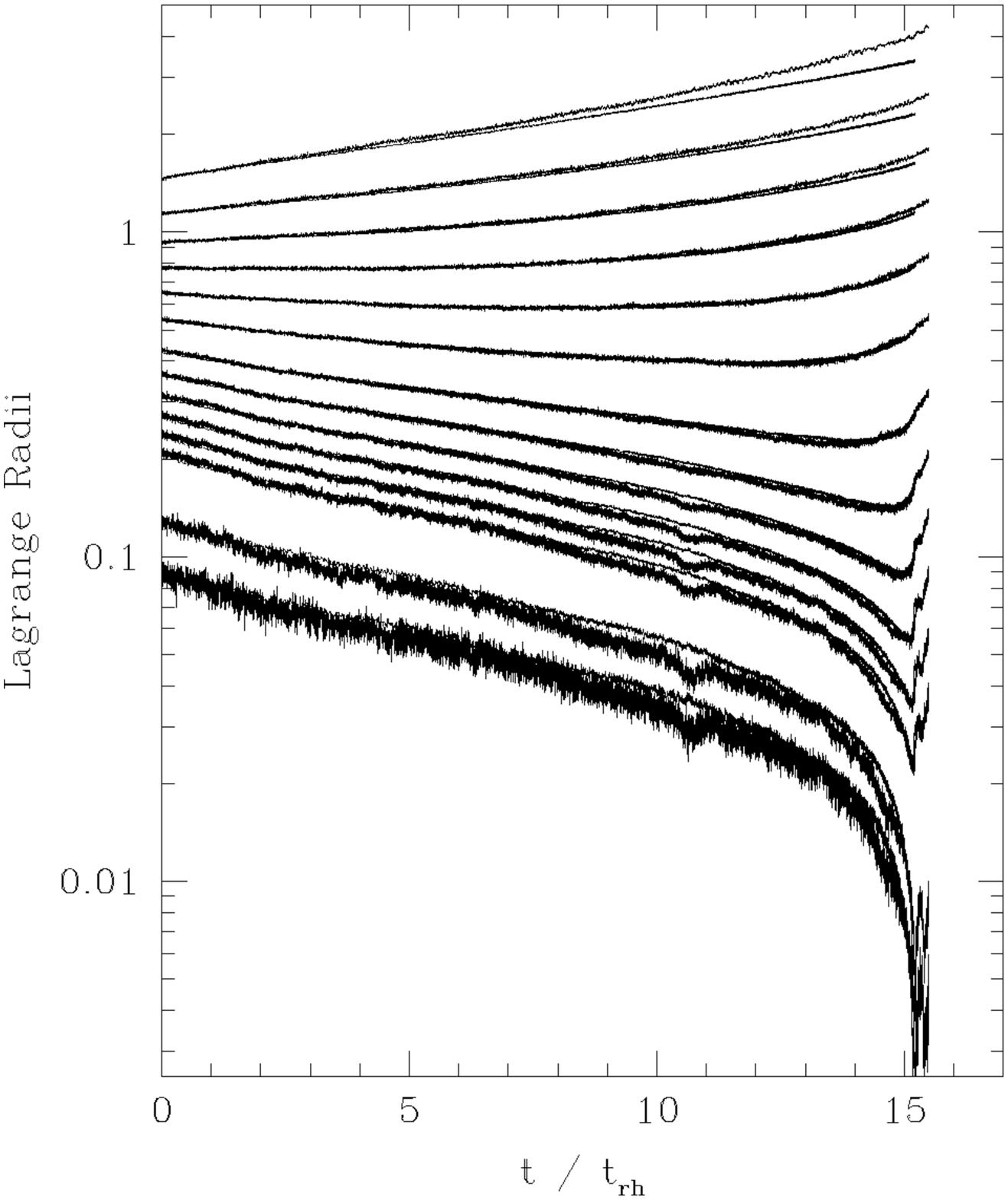}}
  \caption{\it Lagrange radii indicating the evolution of a Plummer
	model globular cluster for an $N$-body simulation and a Monte Carlo
    simulation. The radii correspond to radii containing 0.35, 1, 3.5,
    5, 7, 10, 14, 20, 30, 40, 50, 60, 70, and 80\protect\% of the total
    mass. Figure taken from Joshi {\it et
     al.}\protect~\protect\cite{joshi00}.}
    \label{GC_evolution}
\end{figure}

The core would undergo what is known as core collapse within a few
tens of relaxation times unless these binaries release some of their
binding energy to the cluster. In core collapse, the central density
increases to infinity as the core radius shrinks to zero. An example
of core collapse can be seen in the comparison of two cluster
evolution simulations shown in
Figure~\ref{GC_evolution}~\cite{joshi00}. Note the core collapse when
the inner radius containing 1\% of the total mass dramatically shrinks
after $t \sim 15 \, t_{\rm relax}$. Since these evolution syntheses are
single-mass, Plummer models without binary interactions, the actual
time of core collapse is not representative of a real globular
cluster.

The static description of the structure of globular clusters using
King--Michie or Plummer models provides a framework for describing the
environment of relativistic binaries and their progenitors in globular
clusters. The short-term interactions between stars and degenerate
objects can be analyzed in the presence of this background. Over
longer time scales (comparable to $t_{\rm relax}$), the dynamical
evolution of the distribution function as well as population changes
due to stellar evolution can alter the overall structure of the
globular cluster. We will discuss the dynamical evolution and its
impact on relativistic binaries in
Section~\ref{section:dynamical_evolution}.

Before moving on to the dynamical models and population syntheses of
relativistic binaries, we will first look at the observational
evidence for these objects in globular clusters.

\newpage


\section{Observations}
\label{section:tables}

Observational evidence for relativistic binaries in globular clusters has undergone an explosion in recent years, thanks to concentrated pulsar searches, improved X-ray source positions from Chandra, and optical follow-ups with HST and ground-based telescopes. There are challenges to detecting most binaries since they have generally segregated to the cores of the clusters where crowding can be a problem. Nonetheless, numerous observations of both binaries and their tracer populations have been made in several globular clusters.

\begin{figure}[htbp]
  \def\epsfsize#1#2{0.3#1}
  \centerline{\epsfbox{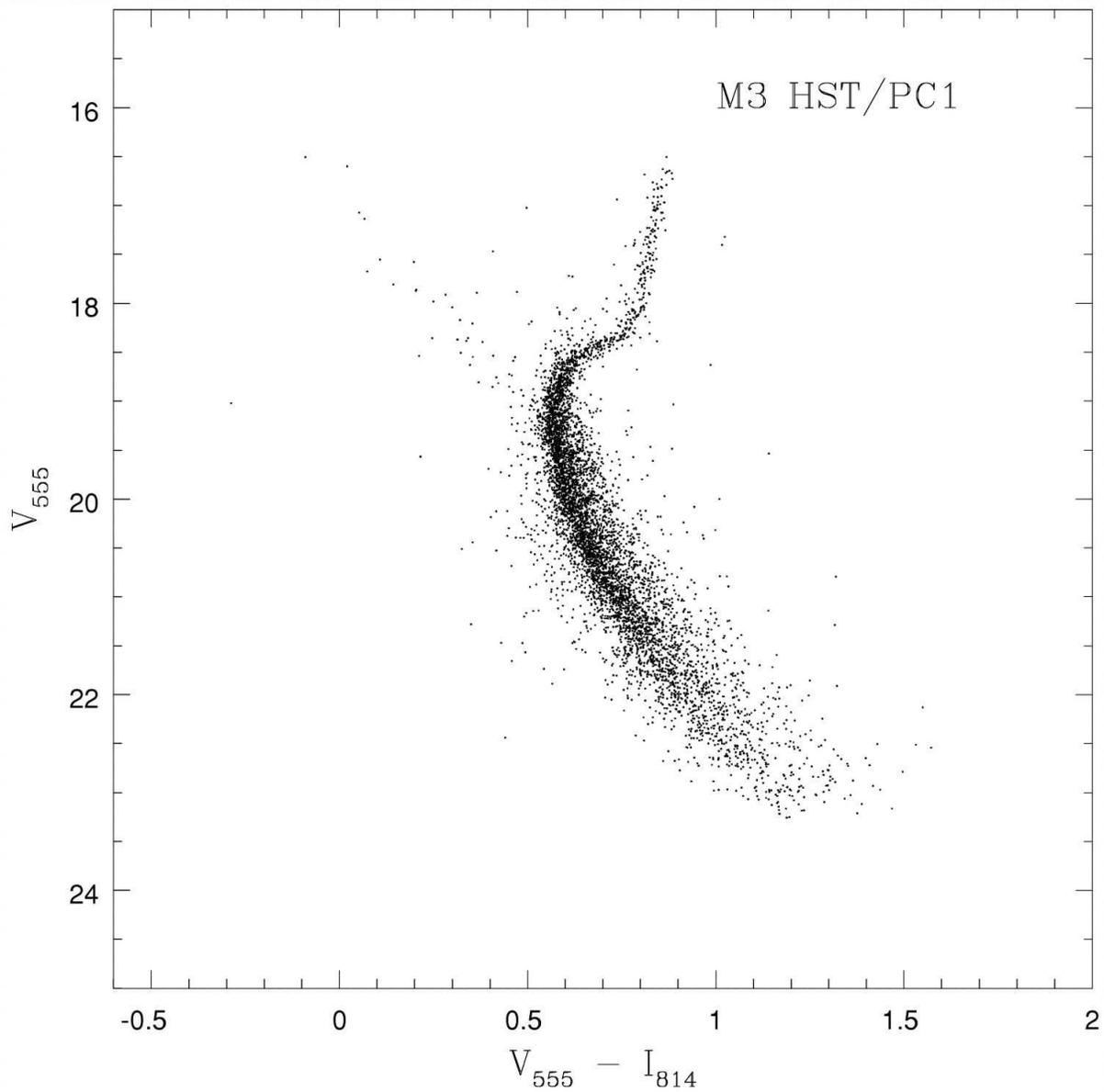}}
  \caption{\it CMD of M3 from the Hubble Space Telescope WFPC2. Note the stars above and to the left of the turn-off. These are the blue stragglers. Figure taken from Zhao and Bailyn \protect~\protect\cite{zhao05}.}
    \label{M3BSS}
\end{figure}

One tracer population of the dynamical processes that may lead to the formation of relativistic binaries is the population of blue stragglers. These are stars that appear on the main sequence above and to the left of the turn-off in the CMD of a globular cluster (see Figure~\ref{M3BSS}). These stars are hot and massive enough that they should have already evolved off the main sequence. Consequently, these objects are thought to arise from stellar coalescences either through the gradual merger of the components of binaries or through direct collisions~\cite{fregeau04,piotto04}. Blue stragglers are some of the most visible and populous evidence of the dynamical interactions that can also give rise to relativistic binaries. For a good description of the use of far-ultraviolet surveys in detecting these objects, see Knigge~\cite{knigge04}. For somewhat older but still valuable reviews on the implications of blue stragglers on the dynamics of globular clusters, see Hut~\cite{hut93} and Bailyn~\cite{bailyn95}.

The globular cluster population of white dwarfs can be used to determine the ages of globular clusters~\cite{moehler04}, and so they have been the focus of targeted searches despite the fact that they are arguably the faintest objects in a globular cluster. These searches have yielded large numbers of globular cluster white dwarfs. For example, a recent search of $\omega$ Centauri has revealed over 2000 white dwarfs~\cite{monelli05}, while Hansen {\it et al.}~\cite{hansen04} have detected 222 white dwarfs in M4. In general, however, these searches uncover single white dwarfs. Optical detection of white dwarfs in binary systems tends to rely on properties of the accretion process related to the binary type. Therefore, searches for cataclysmic variables
generally focus on low-luminosity X-ray sources~\cite{johnston96,
grindlay01a, verbunt01} and on ultraviolet-excess
stars~\cite{grindlay99, knigge02a, margon81}, but these systems are
usually a white dwarf accreting from a low mass star. The class of
``non-flickerers'' which have been detected recently~\cite{cool98,
taylor01} have been explained as He white dwarfs in binaries
containing dark CO white dwarfs~\cite{edmonds99, grindlay01b, hansen03}.

Pulsars, although easily seen in radio, are difficult to detect when
they occur in hard binaries, due to the Doppler shift of the pulse
intervals. Thanks to an improved technique known as an ``acceleration
search''~\cite{middleditch84}, which assumes a constant acceleration
of the pulsar during the observation period, more short orbital period
binary pulsars are being discovered~\cite{camilo00, camilo01,
damico01a, damico01b, freire01b, fruchter00, ransom01}. For a good
review and description of this technique, see
Lorimer~\cite{lorimer01}. The progenitors of the ultracompact
millisecond pulsars are thought to pass through a low-mass X-ray
binary (LMXB) phase~\cite{deutsch00, grindlay01a, rappaport87,
rasio00a}. These systems are very bright and all of them in the
globular cluster system are known. There are, however, several
additional LMXBs that are currently quiescent~\cite{grindlay01a,verbunt04}.

Although there are many theoretical predictions of the existence of black holes in globular clusters (e.g.~\cite{miller02, portegieszwart00b, miller04, dantona04}), there are very few observational hints of them. Measurements of the kinematics of the cores of M15~\cite{gebhardt00,guhathakurta96} and NGC 6752~\cite{drukier03} provide some suggestions of a large, compact mass. However, these observations can also be explained without requiring an intermediate mass black hole~\cite{macnamara03, pasquali04}. The unusual millisecond pulsar in the outskirts of NGC 6752 has also been argued to be the result of a dynamical interaction with a possible binary intermediate mass black hole in the core~\cite{colpi02}. If the velocity dispersion in globular clusters follows the same correlation to black hole mass as in galactic bulges, then there may be black holes with masses in the range $1 - 10^3 M_{\odot}$ in many globular clusters~\cite{zheng01}. Stellar mass black hole binaries may also be visible as low luminosity X-ray sources, but if they are formed in exchange interactions, they will have very low duty cycles and hence are unlikely to be seen~\cite{kalogera04}.

Recent observations and catalogs of known binaries are presented in
the following sections.


\subsection{Cataclysmic variables}
\label{subsection:cataclysmic_variables}

Cataclysmic variables (CVs) are white dwarfs accreting matter from a
companion that is usually a dwarf star or another white dwarf. They have been detected in globular clusters through
identification of the white dwarf itself or through evidence of the
accretion process. White dwarfs managed to avoid detection until
observations with the Hubble Space Telescope revealed
photometric sequences in several globular clusters~\cite{cool96,
cool98, paresce95, renzini96, richer95, richer97, taylor01,hansen04}. Spectral identification of white dwarfs in globular clusters has begun both from the ground with the VLT~\cite{moehler00,moehler04} and in space with the Hubble Space Telescope~\cite{cool98, edmonds99, taylor01,monelli05}. With spectral identification, it will be possible to identify those white dwarfs in hard binaries through Doppler shifts in the ${\rm H}\beta$ line. This approach has promise for detecting a large number of the expected double white dwarf binaries in globular clusters. Photometry has also begun to reveal orbital periods~\cite{neill02,edmonds03b,kaluzny03} of CVs in globular clusters.

Accretion onto the white dwarf may eventually lead to a dwarf nova
outburst. Identifications of globular cluster CVs have been made
through such outbursts in the cores of M5~\cite{margon81}, 47
Tuc~\cite{paresce94}, NGC 6624~\cite{shara96b}, M15~\cite{shara04}, and M22~\cite{anderson03,bond05}. With the exception
of V101 in M5~\cite{margon81}, original searches for dwarf novae
performed with ground based telescopes proved unsuccessful. This is
primarily due to the fact that crowding obscured potential dwarf novae
up to several core radii outside the center of the
cluster~\cite{shara94, shara95a}. Since binaries tend to settle into
the core, it is not surprising that none were found significantly
outside of the core. Subsequent searches using the improved resolution
of the Hubble Space Telescope eventually revealed a few dwarf novae
close to the cores of selected globular clusters~\cite{shara96a,
shara95b, shara96b,shara04,anderson03}.

A more productive approach has been to look for direct evidence of the
accretion around the white dwarf. This can be in the form of excess UV
emission and strong ${\rm H}\alpha$ emission~\cite{ferraro01,
grindlay91, knigge02a, knigge02b, dieball05} from the accretion disk. This technique has
resulted in the discovery of candidate CVs in 47~Tuc~\cite{ferraro01,
knigge02a}, M92~\cite{ferraro00a}, NGC 2808~\cite{dieball05}, NGC 6397~\cite{cool98, edmonds99,
taylor01}, and NGC 6712~\cite{ferraro00b}. The accretion disk can
also be discerned by very soft X-ray emissions. These low luminosity
X-ray binaries are characterized by a luminosity
$L_{\rm X} < 10^{34.5} {\rm\ erg/s}$, which distinguishes them from
the low-mass X-ray binaries with
$L_{\rm X} > 10^{36} {\rm\ erg/s}$. Initial explanations of these
objects focused on accreting white dwarfs~\cite{bailyn91}, and a
significant fraction of them are probably CVs~\cite{verbunt04, webb05}. There have been 10 identified candidate CVs in 6752~\cite{pooley02a}, 19 in 6440~\cite{ pooley02b}, 2 in $\omega$ Cen~\cite{gendre03}, 5 in Terzan 5~\cite{heinke03a}, 22 in 47 Tuc~\cite{edmonds03a}, 5 in M80~\cite{heinke03b}, and 1 in M4~\cite{bassa04}. However, some of the
more energetic sources may be LMXBs in quiescence~\cite{verbunt04}, or even candidate QSO sources~\cite{bassa04}.

The state of the field at this time is one of rapid change as Chandra
results come in and optical counterparts are found for the new X-ray
sources. A living catalog of CVs has been created by Downes {\it et
al.}~\cite{downes01} and may be the best source for confirmed CVs in
globular clusters.


\subsection{Low-mass X-ray binaries}
\label{subsection:lmxbs}

The X-ray luminosities of low-mass X-ray binaries are in the range
$L_{\rm X} \sim 10^{36} - 10^{38} {\rm\ erg/s}$. The upper limit is
close to the Eddington limit for accretion onto a neutron star, so
these systems must contain an accreting neutron star or black
hole. All of the LMXBs in globular clusters contain an accreting
neutron star as they also exhibit X-ray bursts, indicating
thermonuclear flashes on the surface of the neutron
star~\cite{johnston96}. Compared with $\sim 100$ such systems in the
galaxy, there are 13 LMXBs known in globular clusters. The globular
cluster system contains roughly 0.1\% of the mass of the galaxy and
roughly 10\% of the LMXBs. Thus, LMXBs are substantially
over-represented in globular clusters.

Because these systems are so bright in X-rays, the globular cluster
population is completely known -- we expect no new LMXBs to be
discovered in the globular cluster system (unless more multiple
sources are resolved from these 13 sources). The 13 sources are in 12
separate clusters. Three have orbital periods greater than a few
hours, four ultracompact systems have measured orbital periods less
than 1 hour, and six have undetermined orbital periods. The period of X1746-370 in NGC 6441 has recently been measured at $P_{\rm orb} = 5.16~{\rm h}$ using the Rossi X-ray Timing Explorer (RXTE)~\cite{balucinska04}. A member of
the ultracompact group, 4U 1820-30 (X1820-303) in the globular cluster
NGC 6624, has an orbital period of 11 minutes~\cite{stella87}. This is
the shortest known orbital period of any binary and most certainly
indicates a degenerate companion. The orbital period, X-ray
luminosity, and host globular clusters for these systems are given in
Table~\ref{LMXB_Properties}.

\vspace{1 em} \hrule \vspace{0.05 cm} \hrule \vspace{0.2 cm}

\noindent \, \hfill
\begin{minipage}[t]{2.5 cm}
\centering{LMXB Name} \\
\centering{~}
\end{minipage}
\begin{minipage}[t]{2.5 cm}
\centering{Cluster} \\
\centering{~}
\end{minipage}
\begin{minipage}[t]{2.5 cm}
\centering{$ L_{\rm X} $} \\
\centering{($ \times 10^{36} {\rm\ erg/s} $)}
\end{minipage}
\begin{minipage}[t]{1.2 cm}
\centering{$ P_{\rm orb} $} \\
\centering{(hr)}
\end{minipage}
\begin{minipage}[t]{2.5 cm}
\centering{Ref.} \\
\centering{~}
\end{minipage}
\hfill \,

\vspace{0.2 cm} \hrule \vspace{0.2 cm}

\noindent \, \hfill
\begin{minipage}{2.5 cm}~~X0512--401~\end{minipage}
\begin{minipage}{2.5 cm}~~~NGC 1851~\end{minipage}
\begin{minipage}{2.5 cm}\rightline{1.9~~~~~~~}\end{minipage}
\begin{minipage}{1.2 cm}\rightline{$ < 0.85 $~}\end{minipage}
\begin{minipage}{2.5 cm}\rightline{\cite{deutsch00,
sidoli01}}\end{minipage}
\hfill \,

\noindent \, \hfill
\begin{minipage}{2.5 cm}~~X1724--307\footnotemark~\end{minipage}%
\footnotetext{Sidoli {\it et al.}~\cite{sidoli01} give X1724--304.}
\begin{minipage}{2.5 cm}~~~Terzan 2~\end{minipage}
\begin{minipage}{2.5 cm}\rightline{4.3~~~~~~~}\end{minipage}
\begin{minipage}{1.2 cm}\centerline{~~---}\end{minipage}
\begin{minipage}{2.5 cm}\rightline{\cite{deutsch00,
sidoli01}}\end{minipage}
\hfill \,

\noindent \, \hfill
\begin{minipage}{2.5 cm}~~X1730--335~\end{minipage}
\begin{minipage}{2.5 cm}~~~Liller 1~\end{minipage}
\begin{minipage}{2.5 cm}\rightline{2.2~~~~~~~}\end{minipage}
\begin{minipage}{1.2 cm}\centerline{~~---}\end{minipage}
\begin{minipage}{2.5 cm}\rightline{\cite{deutsch00,
sidoli01}}\end{minipage}
\hfill \,

\noindent \, \hfill
\begin{minipage}{2.5 cm}~~X1732--304~\end{minipage}
\begin{minipage}{2.5 cm}~~~Terzan 1~\end{minipage}
\begin{minipage}{2.5 cm}\rightline{0.5~~~~~~~}\end{minipage}
\begin{minipage}{1.2 cm}\centerline{~~---}\end{minipage}
\begin{minipage}{2.5 cm}\rightline{\cite{deutsch00,
sidoli01}}\end{minipage}
\hfill \,

\noindent \, \hfill
\begin{minipage}{2.5 cm}~~X1745--203~\end{minipage}
\begin{minipage}{2.5 cm}~~~NGC 6440~\end{minipage}
\begin{minipage}{2.5 cm}\rightline{0.9~~~~~~~}\end{minipage}
\begin{minipage}{1.2 cm}\centerline{~~---}\end{minipage}
\begin{minipage}{2.5 cm}\rightline{\cite{deutsch00,
sidoli01}}\end{minipage}
\hfill \,

\noindent \, \hfill
\begin{minipage}{2.5 cm}~~X1745--248~\end{minipage}
\begin{minipage}{2.5 cm}~~~Terzan 5~\end{minipage}
\begin{minipage}{2.5 cm}\centerline{~~---}\end{minipage}
\begin{minipage}{1.2 cm}\centerline{~~---}\end{minipage}
\begin{minipage}{2.5 cm}\rightline{\cite{deutsch00}}\end{minipage}
\hfill \,

\noindent \, \hfill
\begin{minipage}{2.5 cm}~~X1746--370~\end{minipage}
\begin{minipage}{2.5 cm}~~~NGC 6441~\end{minipage}
\begin{minipage}{2.5 cm}\rightline{7.6~~~~~~~}\end{minipage}
\begin{minipage}{1.2 cm}\rightline{5.70~}\end{minipage}
\begin{minipage}{2.5 cm}\rightline{\cite{deutsch00, podsiadlowski02,
sidoli01}}\end{minipage}
\hfill \,

\noindent \, \hfill
\begin{minipage}{2.5 cm}~~X1747--313~\end{minipage}
\begin{minipage}{2.5 cm}~~~Terzan 6~\end{minipage}
\begin{minipage}{2.5 cm}\rightline{3.4~~~~~~~}\end{minipage}
\begin{minipage}{1.2 cm}\rightline{12.36~}\end{minipage}
\begin{minipage}{2.5 cm}\rightline{\cite{deutsch00, podsiadlowski02,
sidoli01}}\end{minipage}
\hfill \,

\noindent \, \hfill
\begin{minipage}{2.5 cm}~~X1820--303~\end{minipage}
\begin{minipage}{2.5 cm}~~~NGC 6624~\end{minipage}
\begin{minipage}{2.5 cm}\rightline{40.6~~~~~~~}\end{minipage}
\begin{minipage}{1.2 cm}\rightline{0.19~}\end{minipage}
\begin{minipage}{2.5 cm}\rightline{\cite{deutsch00, podsiadlowski02,
sidoli01}}\end{minipage}
\hfill \,

\noindent \, \hfill
\begin{minipage}{2.5 cm}~~X1832--330~\end{minipage}
\begin{minipage}{2.5 cm}~~~NGC 6652~\end{minipage}
\begin{minipage}{2.5 cm}\rightline{2.2~~~~~~~}\end{minipage}
\begin{minipage}{1.2 cm}\rightline{0.73~}\end{minipage}
\begin{minipage}{2.5 cm}\rightline{\cite{deutsch00,
podsiadlowski02}}\end{minipage}
\hfill \,

\noindent \, \hfill
\begin{minipage}{2.5 cm}~~X1850--087~\end{minipage}
\begin{minipage}{2.5 cm}~~~NGC 6712~\end{minipage}
\begin{minipage}{2.5 cm}\rightline{0.8~~~~~~~}\end{minipage}
\begin{minipage}{1.2 cm}\rightline{0.33~}\end{minipage}
\begin{minipage}{2.5 cm}\rightline{\cite{deutsch00, podsiadlowski02,
sidoli01}}\end{minipage}
\hfill \,

\noindent \, \hfill
\begin{minipage}{2.5 cm}~~X2127+119-1~\end{minipage}
\begin{minipage}{2.5 cm}~~~NGC 7078~\end{minipage}
\begin{minipage}{2.5 cm}\rightline{3.5~~~~~~~}\end{minipage}
\begin{minipage}{1.2 cm}\rightline{17.10~}\end{minipage}
\begin{minipage}{2.5 cm}\rightline{\cite{deutsch00, podsiadlowski02,
sidoli01}}\end{minipage}
\hfill \,

\noindent \, \hfill
\begin{minipage}{2.5 cm}~~X2127+119-2~\end{minipage}
\begin{minipage}{2.5 cm}~~~NGC 7078~\end{minipage}
\begin{minipage}{2.5 cm}\centerline{~~---}\end{minipage}
\begin{minipage}{1.2 cm}\centerline{~~---}\end{minipage}
\begin{minipage}{2.5 cm}\rightline{\cite{deutsch00, podsiadlowski02,nwhite01}}\end{minipage}
\hfill \,

\vspace{0.2 cm} \hrule \vspace{0.05 cm} \hrule \vspace{- 0.5 cm}

\begin{table}[hptb]
  \caption{\it Low-mass X-ray binaries in globular clusters: Host
    clusters and LMXB properties.}
  \label{LMXB_Properties}
\end{table}

The improved resolution of Chandra allows for the possibility of
identifying optical counterparts to LMXBs. If an optical counterpart
can be found, a number of additional properties and constraints for
these objects can be determined through observations in other
wavelengths. In particular, the orbital parameters and the nature of
the secondary can be determined. So far, optical counterparts have
been found for X0512--401 in NGC 1851~\cite{homer01b}, X1745--203 in
NGC 6440~\cite{verbunt00}, X1746--370 in NGC 6441~\cite{deutsch98},
X1830--303 in NGC 6624~\cite{king93}, X1832--330 in NGC
6652~\cite{deutsch00, heinke01}, X1850--087 in NGC
6712~\cite{cudworth88, bailyn88, nieto90}, X1745-248 in Terzan 5~\cite{heinke03a}, and both LMXBs in NGC
7078~\cite{auriere84,nwhite01}. Continued X-ray observations will also further
elucidate the nature of these systems~\cite{mukai00}.

The 13 bright LMXBs are thought to be active members of a larger population of lower luminosity quiescent low mass X-ray binaries (qLMXBs)~\cite{wijnands05}. Early searches performed with ROSAT data (which had a detection limit
of $10^{31} {\rm\ erg/s}$) revealed roughly 30 sources in 19 globular
clusters~\cite{johnston96}. A more recent census of the ROSAT low
luminosity X-ray sources, published by Verbunt~\cite{verbunt01}, lists
26 such sources that are probably related to globular clusters. Recent
observations with the improved angular resolution of Chandra have
begun to uncover numerous low luminosity X-ray candidates for
CVs~\cite{grindlay01a, grindlay01b, heinke01, homer01a, heinke03a, heinke03b, edmonds03a, edmonds03b, gendre03, pooley02a, pooley02b}. For a reasonbly complete discussion of recent observations of qLMXBs in globular clusters, see Verbunt and Lewin~\cite{verbunt04} or Webb and Barret~\cite{webb05} and references therein.


\subsection{Millisecond pulsars}
\label{subsection:millisecond_pulsars}

The population of known millisecond pulsars (MSPs) is one of the fastest growing populations of relativistic binaries in globular clusters. Several ongoing searches are continuing to reveal millisecond pulsars in a number of globular clusters. Previous searches have used deep
multifrequency imaging to estimate the population of pulsars in
globular clusters~\cite{fruchter00}. In this approach, the expected number of pulsars
beaming toward the earth, $N_{\rm puls}$, is determined by the total
radio luminosity observed when the radio beam width is comparable in
diameter to the core of the cluster. If the minimum pulsar luminosity
is $L_{\rm min}$ and the total luminosity observed is $L_{\rm tot}$,
then, with simple assumptions on the neutron star luminosity function,
\begin{equation}
  N_{\rm puls} = \frac{L_{\rm tot}}{L_{\rm min}
  \ln{\left(L_{\rm tot}/L_{\rm min}\right)}}.
  \label{luminosity_number}
\end{equation}
In their observations of 7 globular clusters, Fruchter and Goss have
recovered previously known pulsars in NGC 6440, NGC 6539, NGC 6624,
and 47~Tuc~\cite{fruchter00}. Their estimates based on
Equation~(\ref{luminosity_number}) give evidence of a population of
between 60 and 200 previously unknown pulsars in Terzan~5, and about
15 each in Liller~1 and NGC 6544~\cite{fruchter00}.

Current searches include: Arecibo, which is searching over 22 globular clusters~\cite{hessels04}; Green Bank Telescope (GBT), which is working alone and in conjunction with Arecibo~\cite{jacoby02, hessels04}; the Giant Metrewave Radio Telescope (GMRT), which is searching over about 10 globular clusters~\cite{freire04}; and Parkes, which is searching over 60 globular clusters~\cite{damico03}. Although these searches have been quite successful, they are still subject to certain selection effects such as distance, dispersion measure, and acceleration in compact binaries~\cite{camilo05}. For an excellent review of the properties of all pulsars in globular clusters, see the review by Camilo and Rasio~\cite{camilo05} and references therein. The properties of known pulsars in binary systems with orbital period less than one day are listed in Table~\ref{MSP_Properties}, which is a subset of Table~1 in Camilo and Rasio~\cite{camilo05}.

\vspace{1 em} \hrule \vspace{0.05 cm} \hrule \vspace{0.2 cm}

\noindent \, \hfill
\begin{minipage}[t]{2.5 cm}
\centering{Pulsar} \\
\centering{~}
\end{minipage}
\begin{minipage}[t]{1.5 cm}
\centering{$ P_{\rm spin} $} \\
\centering{(ms)}
\end{minipage}
\begin{minipage}[t]{2.0 cm}
\centering{Cluster} \\
\centering{~}
\end{minipage}
\begin{minipage}[t]{1.2 cm}
\centering{$ P_{\rm orb} $} \\
\centering{(days)}
\end{minipage}
\begin{minipage}[t]{1.6 cm}
\centering{$ e $} \\
\centering{~}
\end{minipage}
\begin{minipage}[t]{1.2 cm}
\centering{$ M_2 $} \\
\centering{($ M_{\odot} $)}
\end{minipage}
\begin{minipage}[t]{1.6 cm}
\centering{Ref.} \\
\centering{~}
\end{minipage}
\hfill \,

\vspace{0.2 cm} \hrule \vspace{0.2 cm}

\noindent \, \hfill
\begin{minipage}{2.5 cm}J0024--7204I~\end{minipage}
\begin{minipage}{1.5 cm}\rightline{3.484~}\end{minipage}
\begin{minipage}{2.0 cm}~47~Tuc~\end{minipage}
\begin{minipage}{1.2 cm}\rightline{0.229~}\end{minipage}
\begin{minipage}{1.6 cm}~$ < $\,0.0004~\end{minipage}
\begin{minipage}{1.2 cm}\rightline{0.013~}\end{minipage}
\begin{minipage}{1.6 cm}\rightline{\cite{freire03}}\end{minipage}
\hfill \,

\noindent \, \hfill
\begin{minipage}{2.5 cm}J0023--7203J~\end{minipage}
\begin{minipage}{1.5 cm}\rightline{2.100~}\end{minipage}
\begin{minipage}{2.0 cm}~47~Tuc\end{minipage}
\begin{minipage}{1.2 cm}\rightline{0.120~}\end{minipage}
\begin{minipage}{1.6 cm}~$ < $\,0.00004~\end{minipage}
\begin{minipage}{1.2 cm}\rightline{0.021~}\end{minipage}
\begin{minipage}{1.6 cm}\rightline{\cite{freire03}}\end{minipage}
\hfill \,

\noindent \, \hfill
\begin{minipage}{2.5 cm}J0024--7204LO~\end{minipage}
\begin{minipage}{1.5 cm}\rightline{2.643~}\end{minipage}
\begin{minipage}{2.0 cm}~47~Tuc\end{minipage}
\begin{minipage}{1.2 cm}\rightline{0.135~}\end{minipage}
\begin{minipage}{1.6 cm}~$ < $\,0.00016~\end{minipage}
\begin{minipage}{1.2 cm}\rightline{0.022~}\end{minipage}
\begin{minipage}{1.6 cm}\rightline{\cite{freire03}}\end{minipage}
\hfill \,

\noindent \, \hfill
\begin{minipage}{2.5 cm}J0024--72P~\end{minipage}
\begin{minipage}{1.5 cm}\rightline{3.643~}\end{minipage}
\begin{minipage}{2.0 cm}~47~Tuc\end{minipage}
\begin{minipage}{1.2 cm}\rightline{0.147~}\end{minipage}
\begin{minipage}{1.6 cm}\rightline{---~}\end{minipage}
\begin{minipage}{1.2 cm}\rightline{0.017~}\end{minipage}
\begin{minipage}{1.6 cm}\rightline{\cite{camilo00}}\end{minipage}
\hfill \,

\noindent \, \hfill
\begin{minipage}{2.5 cm}J0024--72R~\end{minipage}
\begin{minipage}{1.5 cm}\rightline{3.480~}\end{minipage}
\begin{minipage}{2.0 cm}~47~Tuc\end{minipage}
\begin{minipage}{1.2 cm}\rightline{0.066~}\end{minipage}
\begin{minipage}{1.6 cm}\rightline{---~}\end{minipage}
\begin{minipage}{1.2 cm}\rightline{0.026~}\end{minipage}
\begin{minipage}{1.6 cm}\rightline{\cite{camilo00}}\end{minipage}
\hfill \,

\noindent \, \hfill
\begin{minipage}{2.5 cm}J0024--7203U~\end{minipage}
\begin{minipage}{1.5 cm}\rightline{4.342~}\end{minipage}
\begin{minipage}{2.0 cm}~47~Tuc~\end{minipage}
\begin{minipage}{1.2 cm}\rightline{0.429~}\end{minipage}
\begin{minipage}{1.6 cm}~\blankgt 0.000014~\end{minipage}
\begin{minipage}{1.2 cm}\rightline{0.12~}\end{minipage}
\begin{minipage}{1.6 cm}\rightline{\cite{freire03}}\end{minipage}
\hfill \,

\noindent \, \hfill
\begin{minipage}{2.5 cm}J0024--72V~\end{minipage}
\begin{minipage}{1.5 cm}\rightline{4.810~}\end{minipage}
\begin{minipage}{2.0 cm}~47~Tuc\end{minipage}
\begin{minipage}{1.2 cm}\rightline{0.2~}\end{minipage}
\begin{minipage}{1.6 cm}\rightline{---~}\end{minipage}
\begin{minipage}{1.2 cm}\rightline{0.34~}\end{minipage}
\begin{minipage}{1.6 cm}\rightline{\cite{camilo00}}\end{minipage}
\hfill \,

\noindent \, \hfill
\begin{minipage}{2.5 cm}J0024--7204W~\end{minipage}
\begin{minipage}{1.5 cm}\rightline{2.352~}\end{minipage}
\begin{minipage}{2.0 cm}~47~Tuc~\end{minipage}
\begin{minipage}{1.2 cm}\rightline{0.133~}\end{minipage}
\begin{minipage}{1.6 cm}\rightline{---~}\end{minipage}
\begin{minipage}{1.2 cm}\rightline{0.12~}\end{minipage}
\begin{minipage}{1.6 cm}\rightline{\cite{edmonds02}}\end{minipage}
\hfill \,

\noindent \, \hfill
\begin{minipage}{2.5 cm}J0024--72Y~\end{minipage}
\begin{minipage}{1.5 cm}\rightline{2.196~}\end{minipage}
\begin{minipage}{2.0 cm}~47~Tuc\end{minipage}
\begin{minipage}{1.2 cm}\rightline{0.521~}\end{minipage}
\begin{minipage}{1.6 cm}\rightline{---~}\end{minipage}
\begin{minipage}{1.2 cm}\rightline{0.013~}\end{minipage}
\begin{minipage}{1.6 cm}\rightline{\cite{lorimer03}}\end{minipage}
\hfill \,

\noindent \, \hfill
\begin{minipage}{2.5 cm}B1516+02C~\end{minipage}
\begin{minipage}{1.5 cm}\rightline{2.484~}\end{minipage}
\begin{minipage}{2.0 cm}~M5\end{minipage}
\begin{minipage}{1.2 cm}\rightline{0.087~}\end{minipage}
\begin{minipage}{1.6 cm}\rightline{---~}\end{minipage}
\begin{minipage}{1.2 cm}\rightline{0.037~}\end{minipage}
\begin{minipage}{1.6 cm}\rightline{\cite{camilo05}}\end{minipage}
\hfill \,

\noindent \, \hfill
\begin{minipage}{2.5 cm}B1639+36D~\end{minipage}
\begin{minipage}{1.5 cm}\rightline{3.118~}\end{minipage}
\begin{minipage}{2.0 cm}~M13\end{minipage}
\begin{minipage}{1.2 cm}\rightline{0.591~}\end{minipage}
\begin{minipage}{1.6 cm}\rightline{---~}\end{minipage}
\begin{minipage}{1.2 cm}\rightline{0.17~}\end{minipage}
\begin{minipage}{1.6 cm}\rightline{\cite{camilo05}}\end{minipage}
\hfill \,

\noindent \, \hfill
\begin{minipage}{2.5 cm}B1639+36E~\end{minipage}
\begin{minipage}{1.5 cm}\rightline{2.487~}\end{minipage}
\begin{minipage}{2.0 cm}~M13\end{minipage}
\begin{minipage}{1.2 cm}\rightline{0.213~}\end{minipage}
\begin{minipage}{1.6 cm}\rightline{---~}\end{minipage}
\begin{minipage}{1.2 cm}\rightline{0.061~}\end{minipage}
\begin{minipage}{1.6 cm}\rightline{\cite{camilo05}}\end{minipage}
\hfill \,

\noindent \, \hfill
\begin{minipage}{2.5 cm}J1701--3006B~\end{minipage}
\begin{minipage}{1.5 cm}\rightline{3.593~}\end{minipage}
\begin{minipage}{2.0 cm}~M62\end{minipage}
\begin{minipage}{1.2 cm}\rightline{0.144~}\end{minipage}
\begin{minipage}{1.6 cm}~$ < $\,0.00007~\end{minipage}
\begin{minipage}{1.2 cm}\rightline{0.12~}\end{minipage}
\begin{minipage}{1.6 cm}\rightline{\cite{possenti03}}\end{minipage}
\hfill \,

\noindent \, \hfill
\begin{minipage}{2.5 cm}J1701--3006C~\end{minipage}
\begin{minipage}{1.5 cm}\rightline{3.806~}\end{minipage}
\begin{minipage}{2.0 cm}~M62\end{minipage}
\begin{minipage}{1.2 cm}\rightline{0.215~}\end{minipage}
\begin{minipage}{1.6 cm}~$ < $\,0.00006~\end{minipage}
\begin{minipage}{1.2 cm}\rightline{0.069~}\end{minipage}
\begin{minipage}{1.6 cm}\rightline{\cite{possenti03}}\end{minipage}
\hfill \,

\noindent \, \hfill
\begin{minipage}{2.5 cm}J1701--3006E~\end{minipage}
\begin{minipage}{1.5 cm}\rightline{3.234~}\end{minipage}
\begin{minipage}{2.0 cm}~M62\end{minipage}
\begin{minipage}{1.2 cm}\rightline{0.16~}\end{minipage}
\begin{minipage}{1.6 cm}\rightline{---~}\end{minipage}
\begin{minipage}{1.2 cm}\rightline{0.03~}\end{minipage}
\begin{minipage}{1.6 cm}\rightline{\cite{chandler03}}\end{minipage}
\hfill \,

\noindent \, \hfill
\begin{minipage}{2.5 cm}J1701--3006F~\end{minipage}
\begin{minipage}{1.5 cm}\rightline{2.295~}\end{minipage}
\begin{minipage}{2.0 cm}~M62\end{minipage}
\begin{minipage}{1.2 cm}\rightline{0.20~}\end{minipage}
\begin{minipage}{1.6 cm}\rightline{---~}\end{minipage}
\begin{minipage}{1.2 cm}\rightline{0.018~}\end{minipage}
\begin{minipage}{1.6 cm}\rightline{\cite{chandler03}}\end{minipage}
\hfill \,

\noindent \, \hfill
\begin{minipage}{2.5 cm}B1718--19~\end{minipage}
\begin{minipage}{1.5 cm}\rightline{1004.03~}\end{minipage}
\begin{minipage}{2.0 cm}~NGC 6342~\end{minipage}
\begin{minipage}{1.2 cm}\rightline{0.258~}\end{minipage}
\begin{minipage}{1.6 cm}~$ < $\,0.005~\end{minipage}
\begin{minipage}{1.2 cm}\rightline{0.11~}\end{minipage}
\begin{minipage}{1.6 cm}\rightline{\cite{vankerkwijk00}}\end{minipage}
\hfill \,

\noindent \, \hfill
\begin{minipage}{2.5 cm}J1748--2446A~\end{minipage}
\begin{minipage}{1.5 cm}\rightline{11.563~}\end{minipage}
\begin{minipage}{2.0 cm}~Terzan 5~\end{minipage}
\begin{minipage}{1.2 cm}\rightline{0.075~}\end{minipage}
\begin{minipage}{1.6 cm}~$ < $\,0.0012~\end{minipage}
\begin{minipage}{1.2 cm}\rightline{0.087~}\end{minipage}
\begin{minipage}{1.6 cm}\rightline{\cite{lyne00}}\end{minipage}
\hfill \,

\noindent \, \hfill
\begin{minipage}{2.5 cm}J1748--2446AM\end{minipage}
\begin{minipage}{1.5 cm}\rightline{3.569~}\end{minipage}
\begin{minipage}{2.0 cm}~Terzan 5~\end{minipage}
\begin{minipage}{1.2 cm}\rightline{0.443~}\end{minipage}
\begin{minipage}{1.6 cm}\rightline{---~}\end{minipage}
\begin{minipage}{1.2 cm}\rightline{0.13~}\end{minipage}
\begin{minipage}{1.6 cm}\rightline{\cite{ransom05}}\end{minipage}
\hfill \,

\noindent \, \hfill
\begin{minipage}{2.5 cm}J1748--2446N~\end{minipage}
\begin{minipage}{1.5 cm}\rightline{8.666~}\end{minipage}
\begin{minipage}{2.0 cm}~Terzan 5~\end{minipage}
\begin{minipage}{1.2 cm}\rightline{0.385~}\end{minipage}
\begin{minipage}{1.6 cm}~0.000045~\end{minipage}
\begin{minipage}{1.2 cm}\rightline{0.46~}\end{minipage}
\begin{minipage}{1.6 cm}\rightline{\cite{ransom05}}\end{minipage}
\hfill \,

\noindent \, \hfill
\begin{minipage}{2.5 cm}J1748--2446O~\end{minipage}
\begin{minipage}{1.5 cm}\rightline{1.676~}\end{minipage}
\begin{minipage}{2.0 cm}~Terzan 5~\end{minipage}
\begin{minipage}{1.2 cm}\rightline{0.259~}\end{minipage}
\begin{minipage}{1.6 cm}\rightline{---~}\end{minipage}
\begin{minipage}{1.2 cm}\rightline{0.035~}\end{minipage}
\begin{minipage}{1.6 cm}\rightline{\cite{ransom05}}\end{minipage}
\hfill \,

\noindent \, \hfill
\begin{minipage}{2.5 cm}J1748--2446P~\end{minipage}
\begin{minipage}{1.5 cm}\rightline{1.728~}\end{minipage}
\begin{minipage}{2.0 cm}~Terzan 5~\end{minipage}
\begin{minipage}{1.2 cm}\rightline{0.362~}\end{minipage}
\begin{minipage}{1.6 cm}\rightline{---~}\end{minipage}
\begin{minipage}{1.2 cm}\rightline{0.36~}\end{minipage}
\begin{minipage}{1.6 cm}\rightline{\cite{ransom05}}\end{minipage}
\hfill \,

\noindent \, \hfill
\begin{minipage}{2.5 cm}J1748--2446V~\end{minipage}
\begin{minipage}{1.5 cm}\rightline{2.072~}\end{minipage}
\begin{minipage}{2.0 cm}~Terzan 5~\end{minipage}
\begin{minipage}{1.2 cm}\rightline{0.503~}\end{minipage}
\begin{minipage}{1.6 cm}\rightline{---~}\end{minipage}
\begin{minipage}{1.2 cm}\rightline{0.11~}\end{minipage}
\begin{minipage}{1.6 cm}\rightline{\cite{ransom05}}\end{minipage}
\hfill \,

\noindent \, \hfill
\begin{minipage}{2.5 cm}J1807--24A~\end{minipage}
\begin{minipage}{1.5 cm}\rightline{3.059~}\end{minipage}
\begin{minipage}{2.0 cm}~NGC 6544~\end{minipage}
\begin{minipage}{1.2 cm}\rightline{0.071~}\end{minipage}
\begin{minipage}{1.6 cm}\rightline{---~}\end{minipage}
\begin{minipage}{1.2 cm}\rightline{0.0089~}\end{minipage}
\begin{minipage}{1.6 cm}\rightline{\cite{damico01a,
ransom01}}\end{minipage}
\hfill \,

\noindent \, \hfill
\begin{minipage}{2.5 cm}J1911--5958A~\end{minipage}
\begin{minipage}{1.5 cm}\rightline{3.266~}\end{minipage}
\begin{minipage}{2.0 cm}~NGC 6752~\end{minipage}
\begin{minipage}{1.2 cm}\rightline{0.837~}\end{minipage}
\begin{minipage}{1.6 cm}~$ < $\,0.00001~\end{minipage}
\begin{minipage}{1.2 cm}\rightline{0.18~}\end{minipage}
\begin{minipage}{1.6 cm}\rightline{\cite{damico02}}\end{minipage}
\hfill \,

\noindent \, \hfill
\begin{minipage}{2.5 cm}J1911+0102A~\end{minipage}
\begin{minipage}{1.5 cm}\rightline{3.618~}\end{minipage}
\begin{minipage}{2.0 cm}~NGC 6760~\end{minipage}
\begin{minipage}{1.2 cm}\rightline{0.140~}\end{minipage}
\begin{minipage}{1.6 cm}~$ < $\,0.00013~\end{minipage}
\begin{minipage}{1.2 cm}\rightline{0.017~}\end{minipage}
\begin{minipage}{1.6 cm}\rightline{\cite{freire05}}\end{minipage}
\hfill \,

\noindent \, \hfill
\begin{minipage}{2.5 cm}B2127+11C~\end{minipage}
\begin{minipage}{1.5 cm}\rightline{30.529~}\end{minipage}
\begin{minipage}{2.0 cm}~M 15~\end{minipage}
\begin{minipage}{1.2 cm}\rightline{0.335~}\end{minipage}
\begin{minipage}{1.6 cm}~\blankgt 0.681~\end{minipage}
\begin{minipage}{1.2 cm}\rightline{0.92~}\end{minipage}
\begin{minipage}{1.6 cm}\rightline{\cite{anderson90}}\end{minipage}
\hfill \,

\noindent \, \hfill
\begin{minipage}{2.5 cm}J2140--2310A~\end{minipage}
\begin{minipage}{1.5 cm}\rightline{11.019~}\end{minipage}
\begin{minipage}{2.0 cm}~M30~\end{minipage}
\begin{minipage}{1.2 cm}\rightline{0.173~}\end{minipage}
\begin{minipage}{1.6 cm}~$ < $\, 0.00012~\end{minipage}
\begin{minipage}{1.2 cm}\rightline{0.01~}\end{minipage}
\begin{minipage}{1.6 cm}\rightline{\cite{ransom04}}\end{minipage}
\hfill \,

\vspace{0.2 cm} \hrule \vspace{0.05 cm} \hrule \vspace{- 0.5 cm}

\begin{table}[hptb]
  \caption{\it Short orbital period binary millisecond pulsars in
    globular clusters. Host clusters and orbital properties.}
  \label{MSP_Properties}
\end{table}

With the ongoing searches, it can be reasonably expected that the number of millisecond pulsars in binary systems in globular clusters will continue to grow in the coming years.


\subsection{Black holes}
\label{subsection:black_holes}

There have been very few observations of black hole binaries in
globular clusters. Although there have been hints of possible black
hole binaries in extragalactic globular cluster
systems~\cite{angelini01, distefano02}, there are no known black hole
binaries in the galactic globular cluster system. All of the globular
cluster high luminosity LMXBs exhibit the X-ray variability that is indicative of
nuclear burning on the surface of a neutron star. It is possible that some of the recently discovered low luminosity LMXBs may house black holes instead of neutron stars~\cite{verbunt04}, it is more likely that they are simply unusual neutron star LMXBs in quiesence~\cite{wijnands05}. Finally, there is very circumstantial evidence for the possible existence of an intermediate mass black hole (IMBH) binary in NGC 6752 based upon an analysis of the MSP binary PSR A~\cite{colpi03,colpi02}.

\newpage


\section{Relativistic Binaries}
\label{section:relativistic_binaries}

Relativistic binaries are binary systems with at least one degenerate
or collapsed object and an orbital period such that they will be
brought into contact within a Hubble time. (Note that this definition
also includes binaries which are already in contact.) Outside of dense
stellar clusters, most relativistic binary systems arise from
primordial binary systems whose evolution drives them to tight,
ultracompact orbits. The dynamical processes in globular clusters can
drive wide binary systems toward short orbital periods and can also
insert degenerate or collapsed stars into relativistic orbits with
other stars. Before addressing specific evolutionary scenarios, we
will present the generic features of binary evolution that lead to the
formation of relativistic binaries.


\subsection{Binary evolution}
\label{subsection:binary_evolution}

\begin{figure}[htb]
  \def\epsfsize#1#2{0.6#1}
  \centerline{\epsfbox{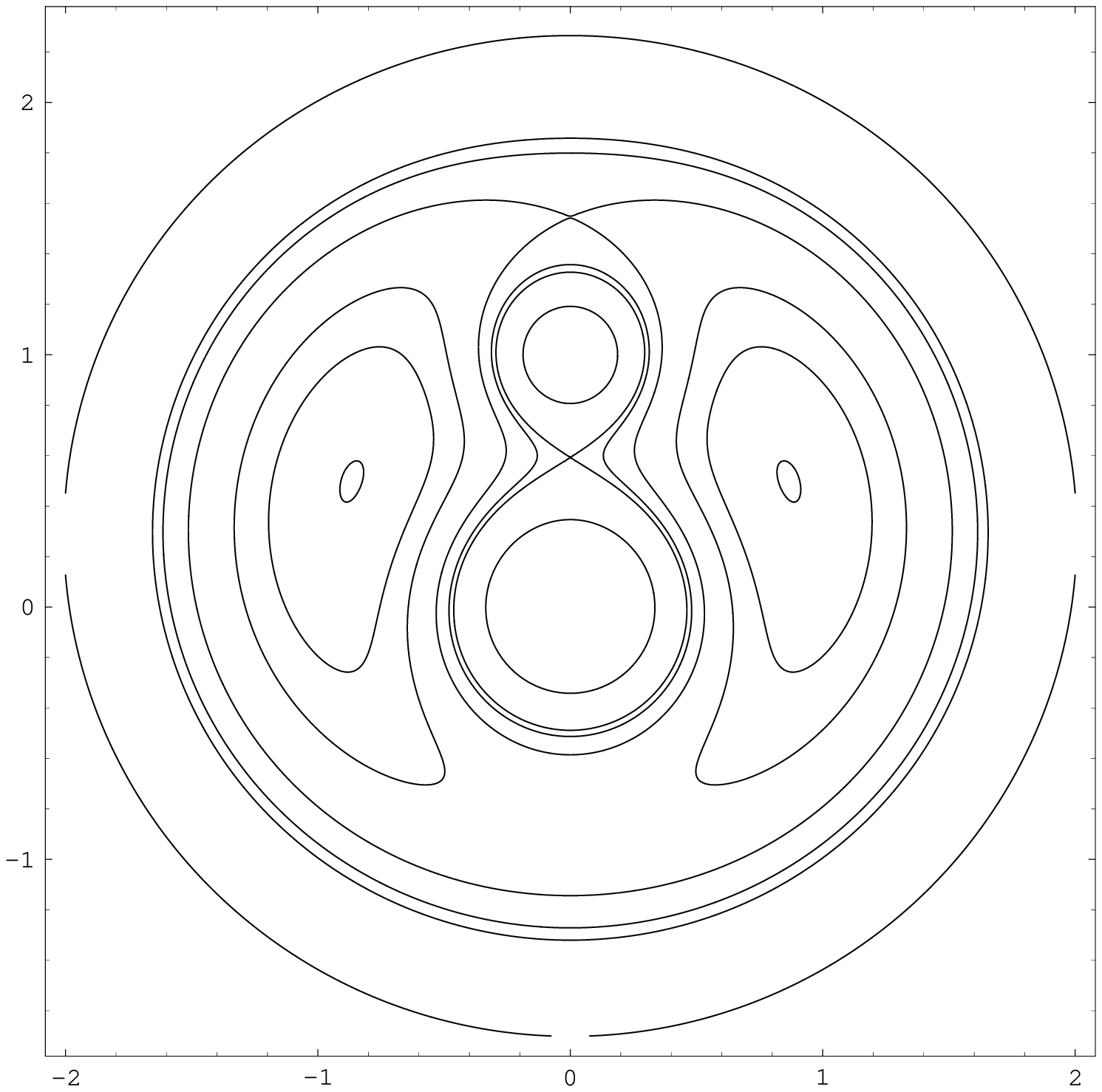}}
  \caption{\it Cross section of equipotential surfaces in the orbital
    plane of a binary with $q = 0.4$. The values of the potential
    surfaces are 5.0, 3.9075, 3.8, 3.559, 3.2, 3.0, and 2.8. The units
    have been normalized to the orbital separation, so $a = 1$.}
  \label{roche_lobe}
\end{figure}

The evolution of a binary system of two main-sequence (MS) stars can 
significantly affect the evolution of both component stars if the
orbital separation is sufficiently small. If the orbital period is less than about 10 days, tidal interactions will have circularized the orbit during the pre- and early main-sequence phase.~\cite{goldman91, zahn89a, zahn89b} Both stars start in the main sequence with the mass of the primary $M_{\rm p}$, and the mass of the secondary $M_{\rm s}$, defined such that $M_{\rm p} \geq M_{\rm s}$. The binary system is described by the orbital separation $a$, and the mass ratio of the components $q \equiv M_{\rm s}/M_{\rm p}$. The gravitational potential of the binary system is described by the Roche model where each star dominates the gravitational potential inside regions called Roche lobes. The two Roche lobes meet at the inner Lagrange point along the line joining the two stars. Figure~\ref{roche_lobe} shows equipotential surfaces in the orbital plane for a binary with $q = 0.4$. If either star fills its Roche lobe, matter will stream from the Roche lobe filling star through the inner Lagrange point to the other star in a process known as Roche lobe overflow (RLOF). This mass transfer affects both the evolution of the components of the binary as well as the binary properties such as orbital period and eccentricity.

Roche lobe overflow can be triggered by the evolution of the binary 
properties or by evolution of the component stars. On the one hand,
the orbital separation of the binary can change so that the Roche lobe
can shrink to within the surface of one of the stars. On the other
hand, stellar evolution may eventually cause one of the stars to
expand to fill its Roche lobe. When both stars in the binary are
main-sequence stars, the latter process is more common. Since the more
massive star will evolve first, it will be the first to expand and
fill its Roche lobe. At this stage, the mass exchange can be
conservative (no mass is lost from the binary) or non-conservative
(mass is lost). Depending on the details of the mass exchange and the
evolutionary stage of the mass-losing star there are several outcomes
that will lead to the formation of a relativistic binary. The primary
star can lose its envelope, revealing its degenerate core as either a
helium, carbon-oxygen, or oxygen-neon white dwarf; it can explode as a
supernova, leaving behind a neutron star or a black hole; or it can
simply lose mass to the secondary so that they change roles. Barring
disruption of the binary, its evolution will then continue. In most
outcomes, the secondary is now the more massive of the two stars and
it may evolve off the main sequence to fill its Roche lobe. The
secondary can then initiate mass transfer or mass loss with the result
that the secondary also can become a white dwarf, neutron star, or
black hole.

The relativistic binaries that result from this process fall into a
number of observable categories. A WD--MS or WD--WD binary may
eventually become a cataclysmic variable once the white dwarf begins
to accrete material from its companion. If the companion is a
main-sequence star, RLOF can be triggered by the evolution of the
companion. If the companion is another white dwarf, then RLOF is
triggered by the gradual shrinking of the orbit through the emission
of gravitational radiation. Some WD--WD cataclysmic variables are also
known as AM CVn stars. If the total mass of the WD--WD binary is above
the Chandrasekhar mass, the system may be a progenitor to a type I
supernova.

The orbit of a NS--MS or NS--WD binary will shrink due to the emission
of gravitational radiation. At the onset of RLOF, the binary will become 
either a low-mass X-ray binary (if the donor star is a WD or MS with
$M \leq 2 M_{\odot}$), or a high-mass X-ray binary (if the donor is a
more massive main-sequence star). These objects may further evolve to
become millisecond pulsars if the NS is spun up during the X-ray
binary phase~\cite{davies98, rasio00a}. A NS--NS binary will remain
virtually invisible unless one of the neutron stars is observable as a
pulsar. A BH--MS or BH--WD binary may also become a low- or high-mass
X-ray binary. If the neutron star is observable as a pulsar, a BH--NS
binary will appear as a binary pulsar. BH--BH binaries will be
invisible unless they accrete matter from the interstellar medium. A
comprehensive table of close binary types that can be observed in
electromagnetic radiation can be found in Hilditch~\cite{hilditch01}.

The type of binary that emerges depends upon the orbital separation
and the masses of the component stars. During the evolution of a
$10 M_{\odot}$ star, the radius will slowly increase by a factor of
about two as the star progresses from zero age main sequence to
terminal age main sequence. The radius will then increase by about
another factor of 50 as the star transitions to the red giant phase,
and an additional factor of 10 during the transition to the red
supergiant phase. These last two increases in size occur very quickly
compared with the slow increase during the main-sequence evolution. Depending upon the orbital separation, the onset of RLOF can occur any time during the evolution of the star. Mass transfer can be divided into three cases related to the timing of the onset of RLOF.

\begin{description}
\item[Case A:] If the orbital separation is small enough (usually a
  few days), the star can fill its Roche lobe during its slow
  expansion through the main-sequence phase while still burning
  hydrogen in its core.
\item[Case B:] If the orbital period is less than about 100 days, but
  longer than a few days, the star will fill its Roche lobe during the
  rapid expansion to a red giant with a helium core. If the helium
  core ignites during this phase and the transfer is interrupted, the
  mass transfer is case BB.
\item[Case C:] If the orbital period is above 100 days, the star can
  evolve to the red supergiant phase before it fills its Roche
  lobe. In this case, the star may have a CO or ONe core.
\end{description}

\begin{figure}[htb]
  \def\epsfsize#1#2{0.5#1}
  \centerline{\epsfbox{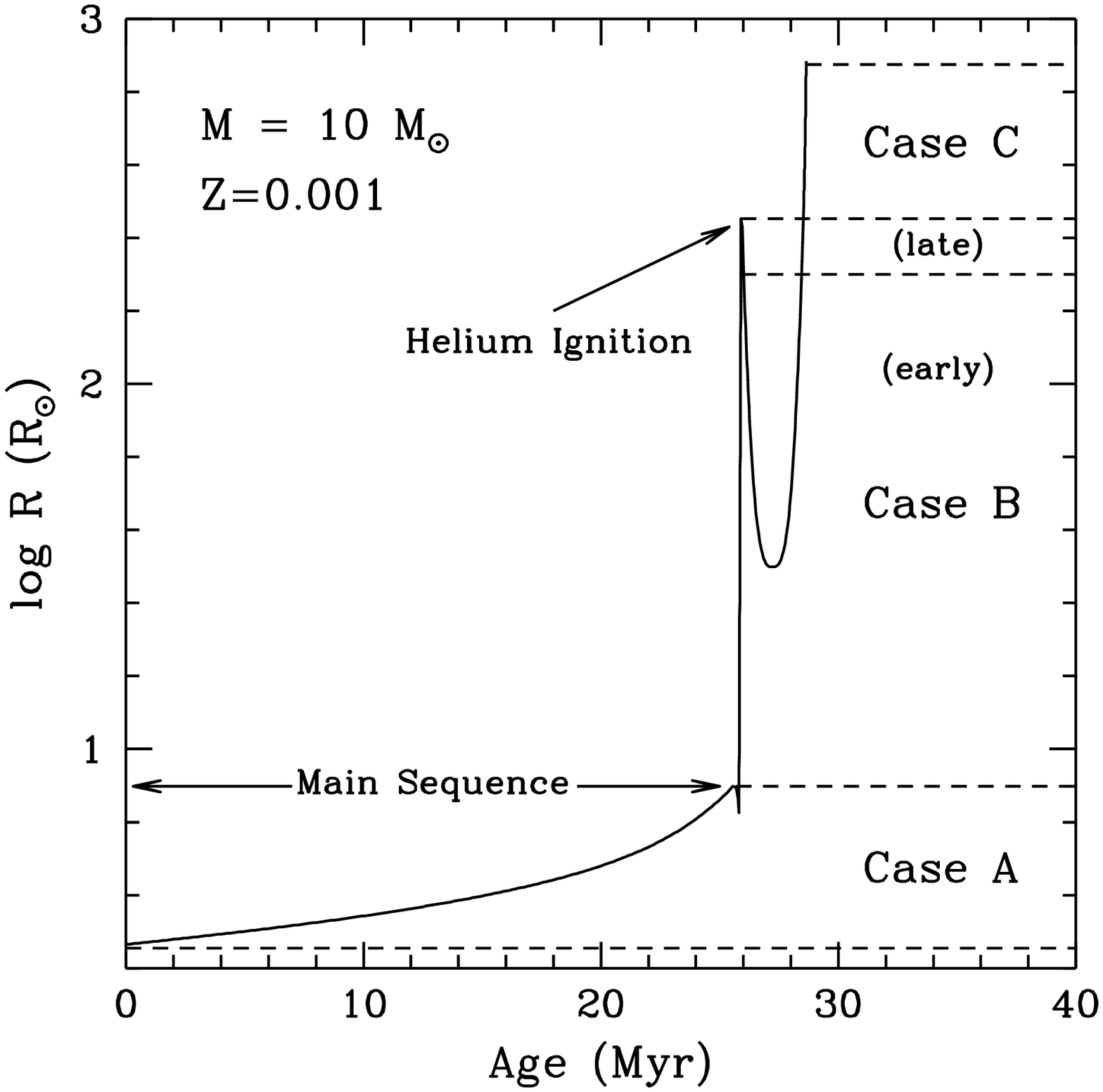}}
  \caption{\it Evolution of the radius for a $10 M_{\odot}$ star with
    $Z = 0.001$. Figure taken from Pfahl {\it et
    al.}\protect~\protect\cite{pfahl02a}.}
  \label{radial_evolution}
\end{figure}

The typical evolution of the radius for a low metallicity star is
shown in Figure~\ref{radial_evolution}. Case A mass transfer occurs
during the slow growth, case B during the first rapid expansion, and
case C during the final expansion phase. The nature of the remnant
depends upon the state of the primary during the onset of RLOF and the
orbital properties of the resultant binary depend upon the details of
the mass transfer.

\subsection{Mass transfer}
\label{subsection:mass_transfer}

Although there are still many unanswered theoretical questions about
the nature of the mass transfer phase, the basic properties of the
evolution of a binary due to mass transfer can easily be
described. The rate at which a star can adjust to changes in its mass
is governed by three time scales. The dynamical time scale results
from the adiabatic response of the star to restore hydrostatic
equilibrium, and can be approximated by the free fall time across the
radius of the star
\begin{equation}
  t_{\rm dyn} \simeq \left(\frac{2 R^3}{G M}\right)^{1/2} \sim
  40\left[ \left( \frac{R}{R_{\odot}} \right)^3
  \frac{M_{\odot}}{M}\right]^{1/2} \!\!\!\!\!\! {\rm\ min},
  \label{dynamical_time}
\end{equation}
where $M$ and $R$ are the mass and radius of the star. The thermal
equilibrium of the star is restored over a longer period given by the
thermal time scale
\begin{equation}
  t_{\rm th} \simeq \frac{G M^2}{R L} \sim 3 \times 10^7 
  \left(\frac{M}{M_{\odot}}\right)^2
  \frac{R_{\odot}}{R}\frac{L_{\odot}}{L} {\rm\ yr},
  \label{thermal_time}
\end{equation}
where $L$ is the luminosity of the star. Finally, the main-sequence
lifetime of the star itself provides a third time scale, which is also
known as the nuclear time scale:
\begin{equation}
  t_{\rm nuc} \sim 7 \times 10^9
  \frac{M}{M_{\odot}}\frac{L_{\odot}}{L} {\rm yr}.
  \label{nuclear_time}
\end{equation}

The rate of mass transfer/loss from the Roche lobe filling star is
governed by how the star's radius changes in response to changes in
its mass. Hjellming and Webbink~\cite{hjellming87} describe these
changes and the response of the Roche lobe to mass changes in the
binary using the radius-mass exponents,
$\zeta \equiv d \ln{R}/d \ln{M}$, for each of the three processes
described in Equations~(\ref{dynamical_time}, \ref{thermal_time},
\ref{nuclear_time}) and defining
\begin{equation}
  \zeta_{\rm L} = (1 + q)\frac{d \ln{R_{\rm L}}}{d \ln{q}}
  \label{roche_lobe_exponent}
\end{equation}
for the Roche lobe radius-mass exponent. If $\zeta_{\rm L} > \zeta_{\rm dyn}$, the star cannot adjust to the Roche lobe, then the mass transfer takes place on a dynamical time scale and is limited only by the rate at which material can stream through the inner Lagrange point. If $\zeta_{\rm dyn} > \zeta_{\rm L} > \zeta_{\rm th}$, then the mass transfer rate is governed by the slow expansion of the star as it relaxes toward thermal equilibrium, and it occurs on a thermal time scale. If both $\zeta_{\rm dyn}$ and $\zeta_{\rm th}$ are greater than $\zeta_{\rm L}$, then the mass loss is driven either by stellar
evolution processes or by the gradual shrinkage of the orbit due to
the emission of gravitational radiation. The time scale for both of
these processes is comparable to the nuclear time scale. A good
analysis of mass transfer in cataclysmic variables can be found in
King {\it et al.}~\cite{king96}.

Conservative mass transfer occurs when there is no mass loss from the
system. During conservative mass transfer, the orbital elements of the
binary can change. Consider a system with total mass $M = M_1 + M_2$
and semi-major axis $a$. The total orbital angular momentum
\begin{equation}
  J = \left[\frac{GM_1^2M_2^2 a}{M}\right]^{1/2}
  \label{orbital_J}
\end{equation}
is a constant, and we can write $a \propto (M_1 M_2)^{-2}$. Using
Kepler's third law and denoting the initial values by a subscript $i$,
we find:
\begin{equation}
  \frac{P}{P_i} = \left[\frac{M_{1i} M_{2i}}{M_1 M_2}\right]^3.
  \label{period_change}
\end{equation}
Differentiating Equation~(\ref{period_change}) and noting that
conservative mass transfer requires $\dot{M}_1 = - \dot{M}_2$ gives:
\begin{equation}
  \frac{\dot{P}}{P} = \frac{3\dot{M}_1\left(M_1-M_2\right)}{M_1M_2}.
  \label{Pdot}
\end{equation}
Note that if the more massive star loses mass, then the orbital period
decreases and the orbit shrinks. If the less massive star is the
donor, then the orbit expands. Usually, the initial phase of RLOF
takes place as the more massive star evolves. As a consequence, the
orbit of the binary will shrink, driving the binary to a more compact
orbit.

In non-conservative mass transfer, both mass and angular momentum can
be removed from the system. There are two basic non-conservative
processes which are important to the formation of relativistic
binaries -- the common-envelope process and the supernova explosion of
one component of the binary. The result of the first process is often
a short-period, circularized binary containing a white dwarf. Although
the most common outcome of the second process is the disruption of the
binary, occasionally this process will result in an eccentric binary
containing a neutron star.

Common envelope scenarios result when one component of the binary
expands so rapidly that the mass transfer is unstable and the
companion becomes engulfed by the donor star. The companion then
ejects the envelope of the donor star. The energy required to eject
the envelope comes from the orbital energy of the binary and thus the
orbit shrinks. The efficiency of this process determines the final
orbital period after the common envelope phase. This is described by
the efficiency parameter
\begin{equation}
  \alpha_{\rm CE} = \frac{\Delta E_{\rm bind}}{\Delta E_{\rm orb}},
  \label{CE_efficiency}
\end{equation}
where $\Delta E_{\rm bind}$ is the binding energy of the mass stripped
from the envelope and $\Delta E_{\rm orb}$ is the change in the
orbital energy of the binary. The result of the process is the exposed
degenerate core of the donor star in a tight, circular orbit with the
companion. This process can result in a double degenerate binary if
the process is repeated twice or if the companion has already evolved
to a white dwarf through some other process. A brief description of
the process is outlined by Webbink~\cite{webbink84}, and a discussion
of the factors involved in determining $\alpha_{\rm CE}$ is presented
in Sandquist {\it et al.}~\cite{sandquist00}.

The effect on a binary of mass loss due to a supernova can be quite
drastic. Following Padmanabhan~\cite{padmanabhan01}, this process is
outlined using the example of a binary in a circular orbit with radius
$a$. Let $v$ be the velocity of one component of the binary relative
to the other component. The initial energy of the binary is given by
\begin{equation}
  E = \frac{1}{2} \left(\frac{M_1 M_2}{M_1 + M_2}\right) v^2 -
  \frac{G M_1 M_2}{a} = - \frac{G M_1 M_2}{2a}.
  \label{initial_energy}
\end{equation}
Following the supernova explosion of $M_1$, the expanding mass shell
will quickly cross the orbit of $M_2$, decreasing the gravitational
force acting on the secondary. The new energy of the binary is then
\begin{equation}
  E^{\prime} = \frac{1}{2} \frac{M_{\rm NS} M_2}{M_{\rm NS} + M_2}
  v^2 - \frac{G M_{\rm NS} M_2}{a},
  \label{final_energy_1}
\end{equation}
where $M_{\rm NS}$ is the mass of the remnant neutron star. We have
assumed here that the passage of the mass shell by the secondary has
negligible effect on its velocity (a safe assumption, see Pfahl
{\it et al.}~\cite{pfahl02a} for a discussion), and that the primary
has received no kick from the supernova (not necessarily a safe
assumption, but see Davies and Hansen~\cite{davies98} or Pfahl {\it et al.}~\cite{pfahl02b} for an
application to globular cluster binaries). Since we have assumed that
the instantaneous velocities of both components have not been
affected, we can replace them by $v^2 = G\left(M_1 + M_2\right)/a$,
and so
\begin{equation}
  E^{\prime} = \frac{G M_{\rm NS} M_2}{2a}\left(\frac{M_1+M_2}
  {M_{\rm NS} + M_2} -2\right).
  \label{final_energy_2}
\end{equation}
Note that the final energy will be positive and the binary will be
disrupted if $M_{\rm NS} < (1/2)(M_1 + M_2)$. This condition occurs
when the mass ejected from the system is greater than half of the
initial total mass:
\begin{equation}
  \Delta M > \frac{1}{2}\left(M_1 + M_2\right),
  \label{deltaM}
\end{equation}
where $\Delta M = M_1 - M_{\rm NS}$. If the binary is not disrupted,
the new orbit becomes eccentric and expands to a new semi-major axis
given by
\begin{equation}
  a^{\prime} = a\left(\frac{M_1 + M_2 - \Delta M}
  {M_1 + M_2 - 2\Delta M}\right),
  \label{final_a}
\end{equation}
and orbital period
\begin{equation}
  P^{\prime} = P\left(\frac{a^{\prime}}{a}\right)^{3/2}
  \left(\frac{2a^{\prime}-a}{a^{\prime}}\right)^{1/2}.
  \label{final_P}
\end{equation}

We have seen that conservative mass transfer can result in a tighter
binary if the more massive star is the donor. Non-conservative mass
transfer can also drive the components of a binary together during a
common envelope phase when mass and angular momentum are lost from the
system. Direct mass loss through a supernova explosion can also alter
the properties of a binary, but this process generally drives the
system toward larger orbital separation and can disrupt the binary
entirely. With this exception, the important result of all of these
processes is the generation of tight binaries with at least one
degenerate object.

The processes discussed so far apply to the generation of
relativistic binaries anywhere. They occur whenever the orbital
separation of a progenitor binary is sufficiently small to allow for
mass transfer or common envelope evolution. Population distributions
for relativistic binaries are derived from an initial mass function, a
distribution in mass ratios, and a distribution in binary
separations. These initial distributions are then fed into models for
binary evolution such as StarTrack~\cite{belczynski02} or SeBa~\cite{portegieszwart96, nelemans01a} in order to determine rates of production of
relativistic binaries. The evolution of the binary is often determined
by the application of some simple operational formulae such as those
described by Tout {\it et al.}~\cite{tout97} or Hurley {\it et al.}~\cite{hurley00}. For example, Hils,
Bender, and Webbink~\cite{hils90} estimated a population of close
white dwarf binaries in the disk of the galaxy using a Salpeter mass
function, a mass ratio distribution strongly peaked at 1, and a
separation distribution that was flat in $\ln(a)$. Other estimates
of relativistic binaries differ mostly by using different
distributions~\cite{belczynski01, iben97, nelemans01a, nelemans01b}.


\subsection{Globular cluster processes}
\label{subsection:globular_cluster_processes}

When the above evolutionary scenarios are played out in the
environment of a globular cluster, additional mechanisms arise that
enhance the production of relativistic binaries. New binary systems
can be formed by dynamical interaction among three or more single
stars or through tidal capture, and the period distribution and
binary components of existing binary systems can be altered by
interactions with other stars in the cluster. We will discuss here the
broad features of these interactions and how they affect the evolution
of binary systems toward relativistic binaries.

The formation of binaries during the dynamical evolution of globular
clusters can occur either through tidal capture or through $N$-body
interactions. Tidal capture occurs when an encounter between two stars
is close enough that significant tides are raised on each. The tides
excite non-radial oscillations in the stars. If the energy absorbed in
these oscillations is great enough to leave the two stars with
negative total energy, then the system will form a binary. This
process was originally thought to be the dominant channel through
which binaries were formed in globular clusters~\cite{binney87,
fabian75}. It is now thought to be quite rare, as detailed
calculations have shown that the final result is more likely to be
coalescence of the two stars~\cite{bailyn95, hut92a,
rasio00a}. Although $N$-body interactions are less likely to occur
than tidally significant two-body interactions, they are now thought
to be the dominant channel for the formation of binaries during the
evolution of a globular cluster. This process, however, is not likely
to produce more than a few binaries during the lifetime of a
cluster~\cite{binney87, padmanabhan01}.

\begin{figure}[htb]
  \def\epsfsize#1#2{1.0#1}
  \centerline{\epsfbox{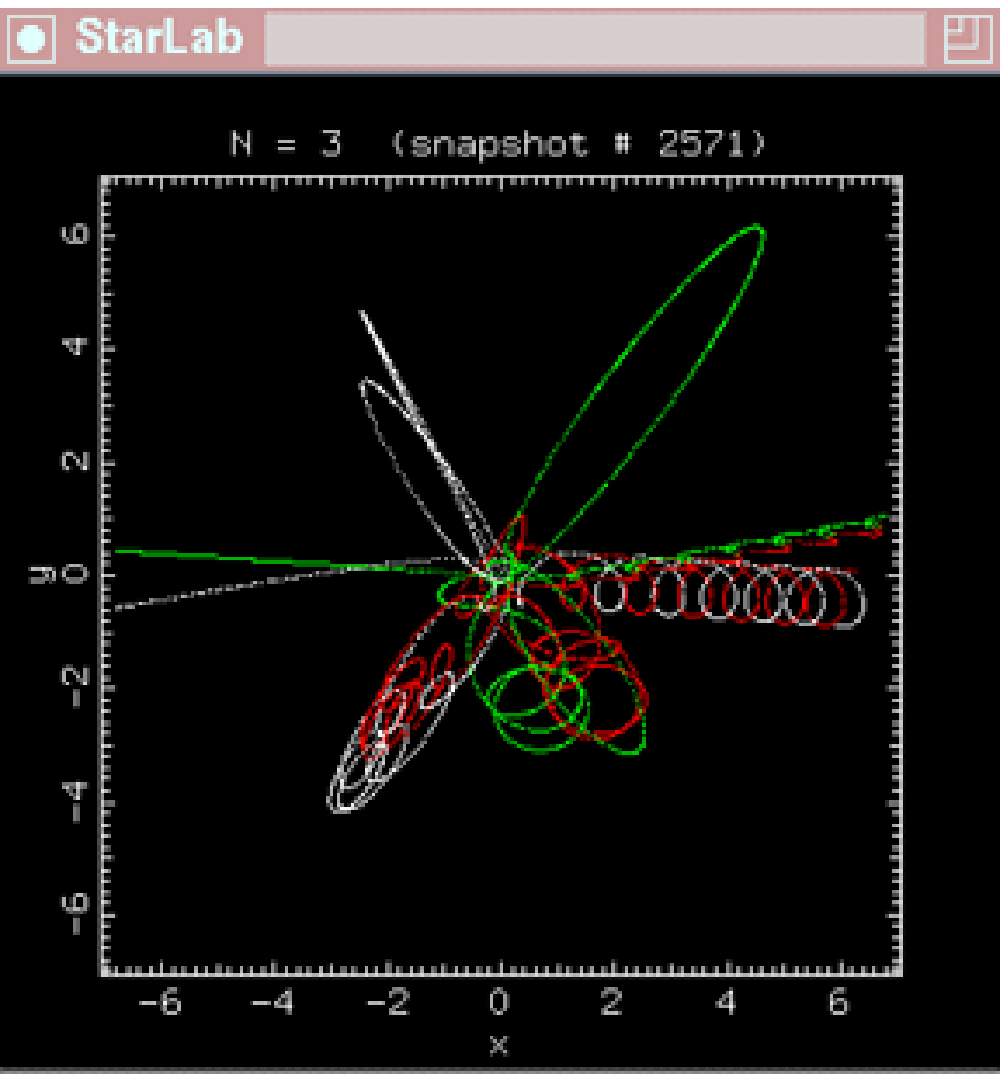}}
  \caption{\it Example of a binary-field star exchange interaction. The
    binary comes in from the right (red-white), while the field star
    (green) enters from the left. After a complicated interaction, the
    white star is ejected and the newly formed red-green binary is in
    a more tightly bound orbit. Figure taken from
    McMillan\protect~\protect\cite{mcmillanweb}.}
  \label{three_body}
\end{figure}

Observations of present binary fractions in globular clusters combined with evolutionary and dynamical simulations indicate initial binary fractions as large as 100\% are not unreasonable~\cite{ivanova05a}. The existence of such a population of primordial binaries provides a much
more efficient channel for the transformation of the initial
distribution in component masses and orbital periods towards higher
mass components and shorter orbital periods. This process follows from
the interaction of primordial binaries with single stars and other
binaries. Three results of the interaction are possible: complete
disruption of the binary, an exchange of energy between the binary and
the field star, or a replacement of one of the binary components by
the field star. When a binary interacts with either a field star or
with another binary, the energy of the interaction is shared among all
stars in the interaction. The result is that the lowest mass object in
the interaction will receive the largest velocity and be more likely
to escape the interaction. In general, these interactions are quite
complex, and must be studied numerically. A typical exchange
interaction between a binary and a field star is shown in
Figure~\ref{three_body}.

If the initial binding energy of the binary is large, the result of
these interactions is to shrink the orbit of the new binary as the
gravitational energy of the binary is used to bring the field star up
to the speeds of the binary components. However, if the binding energy
is low, the field star contributes energy to the components of the
binary, thereby widening the orbit. This is an example of ``Heggie's
Law''~\cite{heggie75}, which can be summarized as {\it hard binaries
get harder and soft binaries get softer}. For roughly equal mass
stars, a binary is considered ``hard'' if its binding energy is
greater than the average kinetic energy of a field star in the cluster
and ``soft'' if its binding energy is less.  For unequal mass
encounters, Hills~\cite{hills90} has shown that the ratio of the
orbital speeds of the binary components to the speed of the impactor
is a better indicator of whether the binding energy will increase or
decrease.

The average kinetic energy of a field star in the cluster is sometimes
related to an effective temperature of the cluster~\cite{heggie75,
mcmillan98, portegieszwart00b} so that
$\langle{m v^2}\rangle = 3kT$. Numerical studies of the outcome of
hard binary interactions indicate that the binding energy of the
binary will increase by about 20\% with each encounter~\cite{hut92b,
  portegieszwart00b}. Since the encounter rate is proportional to the
semi-major axis (or $1/E$) and the energy increase per encounter is
proportional to $E$, the rate of hardening per relaxation time is
independent of the energy and is
$\langle{\Delta E_{\rm bind}}\rangle \sim - 0.6 \,kT/t_{\rm
relax}$~\cite{binney87}.
A common feature of numerical studies of hard binary interactions is
the preferential exchange of high-mass stars and stellar remnants with
the least massive member of the binary~\cite{sigurdsson95}. Thus, the
dynamical interactions in a globular cluster drive the initial orbital
period distribution toward shorter periods by hardening the short
period binaries while disrupting the softer binaries. Through exchange
interactions, the mass distribution of the binary components is also
driven toward higher mass stars, which further enhances the number of
mass-transferring systems that can evolve to become relativistic
binaries. A very useful numerical simulation of multiple star interactions is Fewbody~\cite{fregeau04}.

Because stellar remnants can also be exchanged into hard binaries,
globular cluster evolution opens up a new channel for the formation of
relativistic binaries by introducing evolved components into binary
systems that have not yet undergone a mass transfer phase. A
particularly promising channel involves the exchange of a neutron star
into a binary with a main-sequence star. The binary then undergoes
case B or case C mass transfer with a common envelope phase, resulting
in a NS--WD binary~\cite{rasio00a}. Podsiadlowski {\it et al.}\
describe a similar process without requiring the common envelope
phase~\cite{podsiadlowski02}. Similar interactions can occur to
produce WD--WD binaries if a massive CO or ONe white dwarf is
exchanged into a hard binary. A collaboration of various groups working in stellar dynamics maintains a page at {\tt http://manybody.org/modest/projects.html} that provides a number of useful computational tools for comparing how dynamical interactions can affect different binary evolution codes~\cite{modestweb}.

Black hole binaries can also form as a result of exchange
interactions, but the process is different because black hole
progenitors will evolve so quickly in relation to the relaxation time
of most globular clusters~\cite{kulkarni93, sigurdsson93}. One
scenario that generates black hole binaries in globular clusters is
described by Portegies Zwart and
McMillan~\cite{portegieszwart00b}. Stellar mass black holes of mass
$M \sim 10 M_{\odot}$ will be born early in the life of a globular
cluster and, through mass segregation, they will quickly sink to the
core. Once in the core, these black holes will be so much more massive
than the field stars that they will effectively form their own cluster
and interact solely with themselves. Single black holes will form
binaries with other black holes through three-body encounters; any
black holes which are in binaries with other stars will team up with
another black hole through exchange encounters. This population of
black holes and black hole binaries will then evolve separately from
the rest of the cluster as no other stars will be massive enough to
affect its dynamics.

Current intermediate mass black hole (IMBH) formation scenarios that involve globular clusters can also affect the dynamics of the globular cluster evolution, and therefore, can affect the evolution of binaries within the cluster. In the two most common scenarios, an IMBH is either formed early in the life of the globular cluster through runaway mergers of massive stars~\cite{portegieszwart04b,freitag05,gurkan04} or it is formed through the gradual accumulation of black holes throughout the lifetime of the globular cluster~\cite{miller04}. The existence of an IMBH in a globular cluster can also alter its density profile, and this can have an affect on the rest of the dynamics of the cluster~\cite{baumgardt05}.

We have seen how the dynamics of globular clusters can enhance the
population of progenitors to relativistic binaries, making the
standard channels of mass-transfer more likely to occur. In addition,
globular cluster dynamics can open up new channels for the formation
of relativistic binaries by inserting evolved, stellar remnants such
as neutron stars or white dwarfs into binary systems and by shrinking
the orbits of binary systems to enhance the likelihood of mass
exchange. Finally, binary-single star encounters can simply create
relativistic binaries by inserting two evolved objects into a binary
and then shrinking the orbit to ultracompact periods. We next discuss
the probable rates for the formation of such systems and the dynamical
simulations that are used to synthesize globular cluster populations
of relativistic binaries.

\newpage


\section{Dynamical Evolution}
\label{section:dynamical_evolution}

Simulations of the populations of relativistic binaries in globular
clusters rely on the interplay between the evolution of individual
stars in the progenitor systems and the evolution of globular
clusters. The evolution of stars in the progenitor systems has been
discussed in the previous section and we now turn to techniques for
simulating the evolution of globular clusters.

The evolution of a globular cluster is dominated by the gravitational
interaction between the component stars in the cluster. The overall
structure of the cluster as well as the dynamics of most of the stars
in the cluster are determined by simple $N$-body gravitational
dynamics. However, the evolutionary time scales of stellar evolution
are comparable to the relaxation time and core collapse time of the
cluster. Consequently, stellar evolution affects the masses of the
component stars of the cluster, which affects the dynamical state of
the cluster. Thus, the dynamical evolution of the cluster is coupled
to the evolutionary state of the stars. Also, as we have seen in the
previous section, stellar evolution governs the state of the binary
evolution and binaries may provide a means of support against core
collapse. Thus, the details of binary evolution as coupled with
stellar evolution must also be incorporated into any realistic model
of the dynamical evolution of globular clusters. To close the loop,
the dynamical evolution of the globular cluster affects the
distribution and population of the binary systems in the cluster. In
our case, we are interested in the end products of binary evolution,
which are tied both to stellar evolution and to the dynamical
evolution of the globular cluster. To synthesize the population of
relativistic binaries, we need to look at the dynamical evolution of
the globular cluster as well as the evolution of the binaries in the
cluster. MODEST (MOdeling DEnse STellar systems), a collaboration of various groups working stellar dynamics, maintains a website that provides the latest information about efforts to combine simulations of both the dynamical evolution of $N$-body systems and stellar evolution~\cite{modestweb}.

General approaches to this problem involve solving the $N$-body
problem for the component stars in the cluster and introducing binary
and stellar evolution when appropriate to modify the $N$-body
evolution. There are two fundamental approaches to tackling this
problem -- direct integration of the equations of motion for all $N$
bodies in the system and large-$N$ techniques, such as Fokker--Planck
approximations coupled with Monte Carlo treatments of binaries (see
Heggie {\it et al.}~\cite{heggie98} for a comparison of these
techniques). For a recent review of progress in implementing these techniques, see the summary of the MODEST-2 meeting~\cite{sills03}. In the next two subsections, we discuss the basics of
each approach and their successes and shortfalls. We conclude this
section with a discussion of the recent relativistic binary population
syntheses generated by dynamical simulations.


\subsection{{\it N}-body}
\label{subsection:N-body}

The $N$-body approach to modeling globular cluster dynamics involves
direct calculations of the gravitational interactions between all $N$
bodies in the simulation. In principle this is a very straightforward
approach. The positions of the $N$ objects in the cluster are
determined by direct integration of the $3N$ equations of motion:
\begin{equation}
  {\ddot{\bf r}}_i = -\sum_{j\neq i}{\frac{GM_j({\bf r}_i -
  {\bf r}_j)}{|{\bf r}_j - {\bf r}_i|^3}}.
  \label{equations_of_motion}
\end{equation}
When the positions indicate that objects are sufficiently close to
each other, then the interaction is modeled to determine the
outcome. In order to achieve realistic simulations with tidal
interactions, possible mass transfer, and a mass spectrum of bodies,
detailed stellar evolution and stellar collision models must be
included and calculated.

The two main codes for performing $N$-body simulations are Kira and
NBODYx. The Kira integrator is part of the Starlab environment which
also includes stellar evolution codes and other modules for doing
$N$-body simulations~\cite{starlabweb}. The NBODYx codes have been
developed and improved by Aarseth since the early 1960's. He has two
excellent summmaries of the general properties and development of the
NBODYx codes~\cite{aarseth99b, aarseth99a}. For further details, see Aarseth's book on $N$-body simulations~\cite{aarseth03}. A good summmary of general
$N$-body applications can also be found at the NEMO
website~\cite{nemoweb}. NBODY6++ is a parallelization of the NBODY6 code for use on large computer clusters~\cite{spurzem99}, and a parallel version of Kira is under development~\cite{modestweb}. Most large $N$-body calculations are done with
a special purpose computer called the GRAPE (GRAvity PipE) invented by
Makino~\cite{makino98}. The most recent incarnation of the GRAPE is
the GRAPE 6, which has a theoretical peak speed of 100
Tflops~\cite{grapeweb}. There is also a PCI card version (GRAPE-6A) which is designed for use in PC clusters~\cite{fukushige05}. The GRAPE calculates the accelerations and
jerks for the interaction between each star in the cluster. The next generation GRAPE-DR, which could reach $\sim 1$ Pflops, should be operational in about three years.

The main advantage of $N$-body simulations is the small number of
simplifying assumptions which must be made concerning the dynamical
interactions within the cluster. The specific stars and trajectories
involved in any interactions during the simulation are
known. Therefore, the details for those specific interactions can be
calculated during the simulation. Within the limits of the numerical
errors that accumulate during the calculation~\cite{goodman93}, one
can have great confidence in the results of $N$-body simulations.

Obviously, one of the main computational difficulties is simply the
CPU cost necessary to integrate the equations of motion for $N$
bodies. This scales roughly as $N^3$~\cite{heggie03}. The other
computational difficulty of the direct $N$-body method is the wide
range of precision required~\cite{hut93, heggie03}. Consider the range
of distances, from the size of neutron stars ($\sim 10 {\rm\ km}$) to
the size of the globular cluster
($\sim 50 {\rm\ pc} \sim 10^{15} {\rm\ km}$), spanning 14 orders of
magnitude. If the intent of the calculations is to determine the
frequency of interactions with neutron stars, we have to know the
relative position of every star to within 1 part in $10^{14}$. The
range of time scales is worse yet. Considering that the time for a
close passage of two neutron stars is on the order of milliseconds and
that the age of a globular cluster is $10^{10} {\rm\ yr} \sim 10^{20} {\rm\ ms}$, we find that the time scales span 20 orders of magnitude. These computational requirements coupled with hardware limitations mean that the number of bodies which can be included in a reasonable simulation is no more than $\sim 10^5$. This is about an order of magnitude less than the number of stars in a typical globular cluster.

Although one has great confidence in the results of an $N$-body
simulation, these simulations are generally for systems that are
smaller than globular clusters. Consequently, applications of $N$-body
simulations to globular cluster dynamics involve scaling lower $N$
simulations up to the globular cluster regime. Although many processes
scale with $N$, they do so in different ways. Thus, one scales the
results of an $N$-body simulation based upon the assumption of a
dominant process. However, one can never be certain that the
extrapolation is smooth and that there are no critical points in the
scaling with $N$. One can also scale other quantities in the model, so
that the quantity of interest is correctly
scaled~\cite{portegieszwart99}. An understanding of the nature of the
scaling is crucial to understanding the applicability of $N$-body
simulations to globular cluster dynamics (see
Baumgardt~\cite{baumgardt00} for an example). The scaling problem is
one of the fundamental shortcomings of the $N$-body approach.


\subsection{Fokker--Planck}
\label{subsection:fokker-planck}

The computational limitations of $N$-body simulations can be
sidestepped by describing the system in terms of distribution
functions $f_m(\mathbold{x},\mathbold{v},t)$ with the number of stars
of mass $m$ at time $t$ in the range
$(\mathbold{x},\mathbold{x} + d^3x)$ and
$(\mathbold{v},\mathbold{v} + d^3v)$ given by $dN = f_md^3xd^3v$. This
description requires that either the phase-space element $d^3xd^3v$ be
small enough to be infinitesimal yet large enough to be statistically
meaningful, or that $f_m$ be interpreted as the probability
distribution for finding a star of mass $m$ at a location in phase
space. The evolution of the cluster is then described by the evolution
of $f_m$. The gravitational interaction is provided by a smoothed
gravitational potential $\phi$, which is determined by
\begin{equation}
  \nabla^2\phi = 4 \pi \sum_i{\left[
  m_i\int{f_{m_i}({\bf x}, {\bf v}, t) d^3{\bf v}}\right]}.
  \label{smoothed_potential}
\end{equation}
The effect of gravitational interactions is modeled by a collision
term $\Gamma\left[{f}\right]$ (see~\cite{binney87, padmanabhan01} for
specific descriptions of $\Gamma$). The dynamics of the globular
cluster are then governed by the Fokker--Planck equation:
\begin{equation}
  \frac{\partial f}{\partial t} + {\bf v}\cdot{\bf \nabla}f -
  {\bf \nabla}\phi\cdot\frac{\partial f}{\partial {\bf v}} =
  \Gamma\left[f\right].
  \label{fokker-planck}
\end{equation}

In the Fokker--Planck approach, the mass spectrum of stars is binned,
with a separate $f_m$ for each bin. Increasing the resolution of the
mass spectrum requires increasing the number of distribution functions
and thus increasing the complexity of the problem. Consequently, Fokker--Planck codes can handle at most a few dozen different $f_m$. The inclusion of
additional physical variables such as binaries adds sufficient further
complexity that the codes are taxed beyond their capacity. Methods for numerically solving the Fokker--Planck
equation use either an orbit-averaged form of
Equation~(\ref{fokker-planck})~\cite{cohn79}, or a Monte Carlo
approach~\cite{freitag01, giersz98, giersz01, joshi00, fregeau03}.

The two time scales involved in the evolution of $f_m$ are
$t_{\rm cross}$ (which governs changes in position) and
$t_{\rm relax}$ (which governs changes in energy). The orbit-averaged
form of Equation~(\ref{fokker-planck}) derives from the realization
that changes in position are essentially periodic with orbital period
$T \sim t_{\rm cross} \ll t_{\rm relax}$. Thus, one can average over
the rapid changes in position and retain the slow changes in the phase
space coordinates that occur over relaxation times. Given suitable assumptions on the symmetry of the potential and the velocity distribution, when one does this the Fokker--Planck equation is reduced to an equation involving the energy and the magnitude of the angular momentum. The orbit-averaged solutions of
the Fokker--Planck equation cannot easily handle the effect of
binaries and the binary interactions that occur during the evolution
of a globular cluster~\cite{gao91}. These effects are usually inserted
by hand using statistical methods. The advantages of the
orbit-averaged approach are that one can generalize it to handle
anisotropy in velocity, thus allowing study of the effects of the
galactic gravitational field and tidal stripping. One can also include
the rotation of the cluster~\cite{kim02}.

The more recent Monte Carlo simulations~\cite{giersz98,joshi00,freitag01} do not actually deal with the distribution functions, but rather treat the cluster as a collection of particles that represent a spherical shell of similar stars. Based on the pioneering work of H\'enon~\cite{henon71a,henon71b}, they are able to represent an arbitrary number of species and can follow binary evolution and other effects. The underlying treatment of relaxation throughout the simulation is done in the Fokker-Planck approximation, but the interactions and evolution of the stars are handled on a particle by particle basis. Consequently, these codes are significantly more robust in their ability to handle realistic populations of stars. A nearly continuous mass spectrum can be used and stellar evolution and binarity can be included with relative ease. In addition, both stellar collisions and large-angle scatterings can also be tracked. The primary disadvantages of these Monte Carlo codes are that they require spherical symmetry and that they suffer from statistical noise despite the large number of particles being tracked. For an excellent overview of the implementations and history of the Monte Carlo methods based on H\'enon's work, see Marc Freitag's link on the Working Group 3 page at the MODEST website~\cite{modestweb}.

Another approach to solving the Fokker-Planck equation makes use of the analogy between a globular cluster and a self-gravitating gaseous sphere~\cite{louis91, giersz94}. The most effective use of the gaseous models are in a hybrid code that treats the single stars in a gaseous model while treating the relaxation of binary, three- and four-body interactions using a Monte Carlo code~\cite{giersz00, giersz03}. This approach shows promise for its flexibility in adding new physics.


\subsection{Population syntheses}
\label{subsection:population_syntheses}

Over the last ten years, there have been several works addressing binary populations in
globular clusters~\cite{benacquista99, benacquista01c, davies95a,
davies95b, davies97, davies98, giersz03, hurley03, ivanova05a, miller02, pfahl02a, popov02,
portegieszwart00b, rappaport01, rasio00a, schneider01, shara02, sigurdsson95,
takahashi00}. These have been derived from both dynamical
simulations and static models. Although the motivations have been
varied, it is often possible to extract information about the
resulting populations of relativistic binaries. Despite the differing
models and population synthesis techniques, the predicted populations
are in rough agreement. Here, we summarize the different techniques
and their predictions for relativistic binaries in globular clusters.


\subsubsection{{\it N}-body simulations}
\label{subsubsection:n-body_simulations}

Although $N$-body simulations have the potential to provide the most
detailed population syntheses of relativistic binaries in globular
clusters, there are very few actual populations described in the
literature. Most of the current work that treats binaries in a
consistent and detailed way is limited to open
clusters~\cite{portegieszwart01,hurley01,hurley03,kroupa01b,madsen03} and is focused on
a particular outcome of the binary population, such as blue stragglers~\cite{hurley01}, brown dwarfs~\cite{kroupa01b}, initial binary distributions~\cite{kroupa01c}, or white dwarf CMD sequences~\cite{hurley03}. Portegies Zwart {\it
et al.} focus on photometric observations of open clusters~\cite{portegieszwart01} and on spectroscopy~\cite{portegieszwart04a}. In their comparison of
$N$-body and Fokker--Planck simulations of the evolution of globular
clusters, Takahashi and Portegies Zwart~\cite{takahashi00} followed
the evolution of $N = 1{\rm\ K}, 16{\rm\ K}$, and $32{\rm\ K}$ systems
with initial mass functions given by Equation~(\ref{imf}) and initial
density profiles set up from King models. Although they allowed for
realistic stellar binary evolution in their comparisons, their focus
was on the structural evolution of globular clusters. Consequently
there is no binary population provided. Other $N$-body simulations
suffer from this same problem~\cite{portegieszwart02}. On the other hand, recent work by Shara and Hurley has focused specifically on white dwarf binary populations in globular clusters and has produced a detailed table of close white dwarf binaries that were generated in their simulation~\cite{shara02}.

It is possible to generate a population distribution for black hole
binaries in globular clusters using the $N$-body simulations of
Portegies Zwart and McMillan~\cite{portegieszwart00b} that were
intended to describe the population of black hole binaries that were
{\it ejected} from globular clusters. Their scenario for black hole
binary ejection describes a population of massive stars that evolves
into black holes. The black holes then rapidly segregate to the core
and begin to form binaries. As the black holes are significantly more
massive than the other stars, they effectively form a separate
sub-system, which interacts solely with itself. The black holes form
binaries and then harden through binary-single black hole interactions
that occasionally eject either the binary, the single black hole, or
both.

They simulated this scenario using $N = 2048$ and $N = 4096$ systems
with 1\% massive stars. The results of their simulations roughly
confirm a theoretical argument based on the recoil velocity that a
binary receives during an interaction. Noting that each encounter
increases the binding energy by about 20\% and that roughly $1/3$ of
this energy goes into binary recoil, the minimum binding energy
$E_{\rm b\, min}$ of an ejected black hole binary is
\begin{equation}
  E_{\rm b\,min} \sim 36 W_0 \frac{M_{\rm bh}}{\langle M\rangle} kT,
  \label{binding_energy}
\end{equation}
where $\langle{M}\rangle$ is the average mass of a globular cluster
star and $W_0 = \langle M\rangle |\phi_0|/kT$ is the dimensionless
central potential. After most binaries are ejected,
$\langle{M}\rangle \sim 0.4 M_{\odot}$. After a few gigayears, nearly
all of the black holes were ejected.

At the end of this phase of black hole binary ejection, there is a
50\% chance that a binary remains in the cluster with no other black
hole to eject it. Thus, there should be a stellar mass black hole
binary remaining in about half of the galactic globular clusters. The
maximum binding energy of the remaining black hole binary is
$E_{\rm b\,min}$ and is also given by
Equation~(\ref{binding_energy}). We can then approximate the
distribution in energies of the remaining black hole binaries as being
flat in $\log (E_{\rm b})$. The eccentricities of this population will
follow a thermal distribution with $P(e) = 2e$.


\subsubsection{Monte Carlo simulations}
\label{subsubsection:monte-carlo_simulations}

\begin{figure}[htb]
  \def\epsfsize#1#2{0.5#1}
  \centerline{\epsfbox{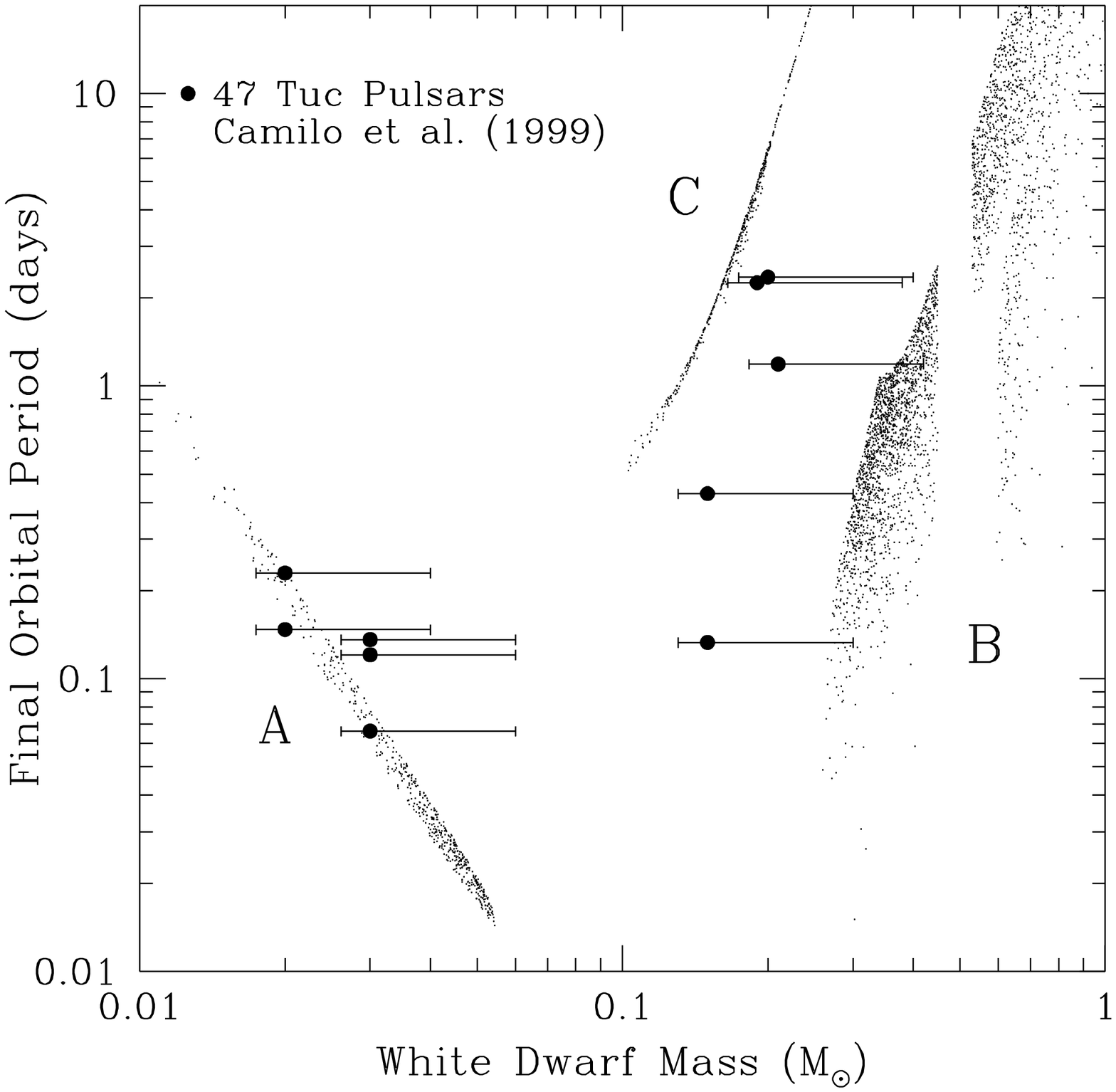}}
  \caption{\it Results of the Monte Carlo simulation of NS--WD binary
    generation and evolution in 47\protect~Tuc. Each small dot
	represents a binary system. The circles and error bars are the 10
	binary pulsars in 47\protect~Tuc with well measured orbits. Systems 
	in A have evolved through mass transfer from the WD to the NS.
	Systems in B have not yet evolved through gravitational radiation to 
	begin RLOF from the WD to the NS. Systems in C will not undergo a 
	common envelope phase. Figure taken from Rasio {\it et 
	al.}\protect~\protect\cite{rasio00a}.}
  \label{WD-NS_in_47Tuc}
\end{figure}

Dynamical Monte Carlo simulations can be used to study the evolution
of binary populations within evolving globular cluster models. Rasio
{\it et al.}~\cite{rasio00a} have used a Monte Carlo approach
(described in Joshi {\it et al.}~\cite{joshi01, joshi00}) to study the
formation and evolution of NS--WD binaries, which may be progenitors of
the large population of millisecond pulsars being discovered in
globular clusters (see
Section~\ref{subsection:millisecond_pulsars}). In addition to
producing the appropriate population of binary millisecond pulsars to
match observations, the simulations also indicate the existence of a
population of NS--WD binaries (see Figure~\ref{WD-NS_in_47Tuc}).

The tail end of the systems in group B of Figure~\ref{WD-NS_in_47Tuc}
represents the NS--WD binaries that are in very short period orbits and
are undergoing a slow inspiral due to gravitational radiation. These
few binaries can be used to infer an order of magnitude estimate on
the population of such objects in the galactic globular cluster
system. If we consider that there are two binaries with orbital period
less than 2000~s out of $\sim 10^6 M_{\odot}$ in 47~Tuc, and assume
that this rate is consistent throughout the globular cluster system as
a whole, we find a total of $\sim 60$ such binaries. Although this
estimate is quite crude, it compares favorably with estimates arrived
at through the encounter rate population syntheses, which are
discussed in Section~\ref{subsubsection:encounter_rates}.

\begin{figure}[htb]
  \def\epsfsize#1#2{0.5#1}
  \centerline{\epsfbox{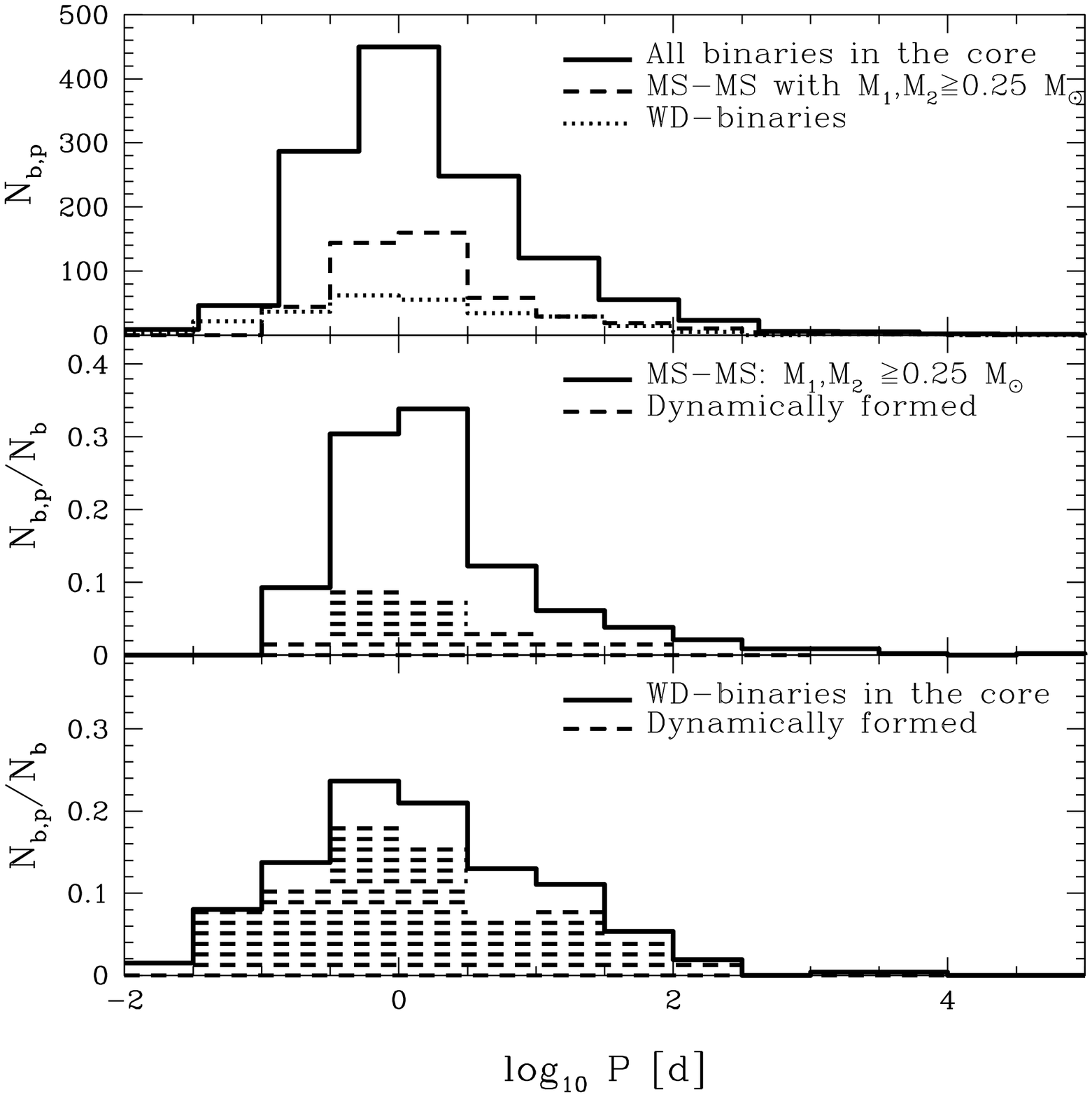}}
  \caption{\it Binary period distributions from the Monte Carlo simulation of binary
    fraction evolution in 47\protect~Tuc. The bottom panel indicates the period distribution for binaries containing at least one white dwarf. $N_{\rm b}$ is the total number of binaries and $N_{\rm b,p}$ is the number of binaries per bin. Figure taken from Ivanova {\it et 
	al.}\protect~\protect\cite{ivanova05a}.}
  \label{ivanova47tuc}
\end{figure}

More recent applications of the Monte Carlo simulations that have focused on the properties of binaries include Fregeau {\it et al.}~\cite{fregeau04} who look at the production of blue stragglers and other collision products as a result of binary interactions in globular clusters and Ivanova {\it et al.}~\cite{ivanova05a} who have studied the evolution of binary fractions in globular clusters. The latter work demonstrates the gradual burning of binaries in the core that delays the collapse of the core. In addition, they have also shown the build-up of short period white dwarf binaries in the core through dynamical interactions (see Figure~\ref{ivanova47tuc}).

There is also great promise for the hybrid gas/Monte Carlo method
being developed by Spurzem and Giersz~\cite{spurzem96}. Their recent
simulation of the evolution of a cluster of 300,000 equal point-mass
stars and 30,000 binaries yields a wealth of detail about the position
and energy distribution of binaries in the
cluster~\cite{giersz00}. Further improvements on their code have resulted in direct integration of the binary-binary and binary-single interactions~\cite{giersz03}. As a result, they have been able to produce empirical cross-sections for eccentricity variations during interactions.


\subsubsection{Encounter rate techniques}
\label{subsubsection:encounter_rates}

One method for exploring the production of relativistic binary populations in globular clusters involves determining the encounter rate expected between different classes of objects in a globular cluster. Sigurdsson and Phinney~\cite{sigurdsson95} use Monte Carlo
simulations of binary encounters to infer populations using a static
background cluster described by an isotropic King--Michie model. Their
results are focused toward predicting the observable end products of
binary evolution such as millisecond pulsars, cataclysmic variables,
and blue stragglers. Therefore, there are no clear descriptions of
relativistic binary populations provided. The work of Ivanova {\it et al.}~\cite{ivanova05a} also uses this technique to determine the evolution of binary fractions, but they also do not provide sufficient detail of the population to distinguish the relativistic binaries from other binaries in the simulation. There is promise to produce a more detailed description of ultracompact X-ray binaries consisting of a white dwarf and a neutron star using encounter rates~\cite{ivanova05b}.

Davies and collaborators use the technique of calculating encounter rates (based on
calculations of cross-sections for various binary interactions and
number densities of stars using King--Michie static models) to
determine the production of end products of binary
evolution~\cite{davies95a, davies95b}. Although they also do not
provide a clear description of a population of relativistic binaries,
their results allow the estimation of such a population.

Using the encounter rates of Davies and collaborators~\cite{davies95a,
davies95b}, one can follow the evolution of binaries injected into
the core of a cluster. A fraction of these binaries will evolve into
compact binaries which will then be brought into contact through the
emission of gravitational radiation. By following the evolution of
these binaries from their emergence from common envelope to contact,
we can construct a population and period distribution for present day
globular clusters~\cite{benacquista99}. For a globular cluster with
dimensionless central potential $W_0 = 12$, Davies~\cite{davies95b}
followed the evolution of 1000 binaries over two runs. The binaries
were chosen from a Salpeter IMF with exponent $\alpha = 2.35$, and the
common envelope evolution used an efficiency parameter
$\alpha_{\rm CE} = 0.4$. One run was terminated after 15~Gyr and the
population of relativistic binaries which had been brought into
contact through gravitational radiation emission was noted. The second
run was allowed to continue until all binaries were either in merged
or contact systems. There are four classes of relativistic binaries
that are brought into contact by gravitational radiation: low mass
white dwarf--white dwarf binaries (${\rm WD}^2_a$) with total mass below
the Chandrasekhar mass; high mass white dwarf--white dwarf binaries
(${\rm WD}^2_b$) with total mass above the Chandrasekhar mass; neutron
star-white dwarf binaries (${\rm NW}$); and neutron star-neutron star
binaries ${\rm NS}^2$. The number of systems brought into contact at the
end of each run is given in Table~\ref{contact_numbers}.

\begin{table}[h]
  \begin{center}
    \begin{tabular}{|c|c|c|c|c|}
      \hline
      $T_{\rm evol}$ & ${\rm WD}_a^2$ & ${\rm WD}_b^2$ & ${\rm NW}$ &
      ${\rm NS}^2$\rule{0 em}{1 em} \\
      \hline
      15 Gyr & 11 & 0 & 10 & \blankzero 1 \\
      $\infty$ & 57 & 0 & 74 & 18 \\
      \hline
    \end{tabular}
  \end{center}
  \caption{\it Number of relativistic binaries brought into contact
    through binary interactions.}
  \label{contact_numbers}
\end{table}

In the second run, the relativistic binaries had all been brought into
contact. In similar runs, this occurs after another 15~Gyr. An
estimate of the present-day period distribution can be made by
assuming a constant merger rate over the second 15~Gyr. Consider the
total number of binaries that will merge to be described by
$n(t)$. Thus, the merger rate is $\eta = -dn/dt$. Assuming that the
mergers are driven solely by gravitational radiation, we can relate
$n(t)$ to the present-day period distribution. We define $n(P)$ to be
the number of binaries with period less than P, and thus
\begin{equation}
  \eta = -\frac{dn}{dt} = -\frac{dn}{dP}\frac{dP}{dt},
  \label{merger_rate}
\end{equation}
so
\begin{equation}
  \frac{dn}{dP} = \frac{-\eta}{dP/dt}.
  \label{period_distribution}
\end{equation}

The merger rate is given by the number of mergers of each binary type
per 1000 primordial binaries per 15~Gyr. If the orbits have been
circularized (which is quite likely if the binaries have been formed
through a common envelope), the evolution of the period due to
gravitational radiation losses is given by~\cite{hils90}
\begin{equation}
  \frac{dP}{dt} = -k_0P^{-5/3},
  \label{gr_period_evolution}
\end{equation}
where $k_0$ is given by
\begin{equation}
  k_0 = \frac{96}{5} (2\pi)^{8/3}\frac{G^{5/3}}{c^5}{\cal M}^{5/3},
  \label{knaught}
\end{equation}
with the ``chirp mass'' ${\cal M}^{5/3} \equiv M_1M_2(M_1 +M_2)^{-1/3}$. 

Following this reasoning and using the numbers in
Table~\ref{contact_numbers}, we can determine the present day
population of relativistic binaries per 1000 primordial binaries. To
find the population for a typical cluster, we need to determine the
primordial binary fraction for globular clusters. Estimates of the
binary fraction in globular clusters range from 13\% up to about 40\%
based on observations of either eclipsing binaries~\cite{albrow01,
yan94, yan96} or luminosity functions~\cite{rubenstein97,
rubenstein99}. Assuming a binary fraction of 30\%, we can determine
the number of relativistic binaries with short orbital period
$(P_{\rm orb} < P_{\rm max})$ for a typical cluster with
$10^6 M_{\odot}$ and the galactic globular cluster system with
$10^{7.5} M_{\odot}$~\cite{sigurdsson95} by simply integrating the
period distribution from contact $P_{\rm c}$ up to $P_{\rm max}$
\begin{equation}
  N = \int_{P_{\rm c}}^{P_{\rm max}}{\frac{\eta}{k_0}P^{5/3}dP}.
  \label{encounter_number}
\end{equation}
The value of $P_{\rm c}$ can be determined by using the Roche lobe
radius of Eggleton~\cite{eggleton83},
\begin{equation}
  R_{\rm L} = \frac{0.49 q^{2/3}}{0.6 q^{2/3} +
  \ln{\left(1 + q^{1/3}\right)}}a,
  \label{eggleton_roche_lobe}
\end{equation}
and stellar radii as determined by Lynden-Bell and
O'Dwyer~\cite{lyndenbell01}.

The expected populations for an individual cluster and the galactic
cluster system are shown in Table~\ref{encounter_populations} using
neutron star masses of $1.4 M_{\odot}$, white dwarf masses of $0.6
M_{\odot}$ and $0.3 M_{\odot}$, and $P_{\rm max} = 2000 {\rm\ s}$.

\begin{table}[h]
  \begin{center}
    \begin{tabular}{|c|c|c|c|}
      \hline
      Object & ${\rm WD}_a^2$ & ${\rm NW}$ &
      ${\rm NS}^2$\rule{0 em}{1 em} \\
      \hline
      Cluster & \blankzero \blankzero 5.6 & \blankzero \blankzero 4.0 &
0.5 \\
      System  & 176.5 & 125.2 & \blankdot 16 \\
      \hline
    \end{tabular}
  \end{center}
  \caption{\it Encounter rate estimates of the population of
    relativistic binaries in a typical globular cluster and the
    galactic globular cluster system.}
  \label{encounter_populations}
\end{table}

Although we have assumed the orbits of these binaries will be
circularized, there is the possible exception of ${\rm NS}^2$
binaries, which may have a thermal distribution of eccentricities if
they have been formed through exchange interactions rather than
through a common envelope. In this case,
Equations~(\ref{gr_period_evolution}) and~(\ref{knaught}) are no
longer valid. An integration over both period and eccentricity, using
the formulae of Pierro and Pinto~\cite{pierro96}, would be required.


\subsubsection{Semi-empirical methods}
\label{subsubsection:semi-empirical_methods}

The small number of observed relativistic binaries can be used to
infer the population of dark progenitor
systems~\cite{benacquista01c}. For example, the low-mass X-ray binary
systems are bright enough that we see essentially all of those that
are in the galactic globular cluster system. If we assume that the
ultracompact ones originate from detached WD--NS systems, then we can
estimate the number of progenitor systems by looking at the time spent
by the system in both phases. Let $N_{\rm X}$ be the number of
ultracompact LMXBs and $T_{\rm X}$ be their typical lifetime. Also,
let $N_{\rm det}$ be the number of detached WD--NS systems that will
evolve to become LMXBs, and $T_{\rm det}$ be the time spent during the
inspiral due to the emission of gravitational radiation until the
companion white dwarf fills its Roche lobe. If the process is
stationary, we must have
\begin{equation}
  \frac{N_{\rm X}}{T_{\rm X}} = \frac{N_{\rm det}}{T_{\rm det}}.
  \label{population_ratio}
\end{equation}
The time spent in the inspiral phase can be found from integrating
Equation~(\ref{gr_period_evolution}) to get
\begin{equation}
  T_{\rm det} = \frac{3}{8k_0}\left(P_0^{8/3}-P_{\rm c}^{8/3}\right)
  \label{detached_time}
\end{equation}
where $P_0$ is the period at which the progenitor emerges from the
common envelope and $P_{\rm c}$ is the period at which RLOF from the
white dwarf to the neutron star begins. Thus, the number of detached
progenitors can be estimated from
\begin{equation}
  N_{\rm det} = \frac{N_{\rm X}}{T_{\rm X}}
  \frac{3}{8k_0}\left(P_0^{8/3}-P_{\rm c}^{8/3}\right).
  \label{detached_number}
\end{equation}

There are four known ultracompact LMXBs~\cite{deutsch00} with orbital 
periods small enough to require a degenerate white dwarf companion to
the neutron star. There are six other LMXBs with unknown orbital
periods. Thus, $4 \leq N_{\rm X} \leq 10$. The lifetime $T_{\rm X}$
is rather uncertain, depending upon the nature of the mass transfer
and the timing when the mass transfer would cease. A standard
treatment of mass transfer driven by gravitational radiation alone
gives an upper bound of $T_{\rm X} \sim 10^9 {\rm\ yr}$~\cite{rappaport87}, but other effects such as tidal heating or irradiation may shorten this to $T_{\rm X} \sim 10^7 {\rm\ yr}$~\cite{applegate94,rasio00a}. The value of $P_0$ depends critically upon the evolution of the neutron star--main-sequence binary, and is very uncertain. Both $k_0$ and $P_{\rm c}$ depend upon the masses of the WD secondary and the NS primary. For a rough estimate, we take the mass of the secondary to be a typical He WD of mass $0.4 M_{\odot}$ and the mass of the primary to be $1.4 M_{\odot}$. Rather than estimate the typical period of emergence from the common envelope, we arbitrarily choose $P_0 = 2000 {\rm\ s}$. We can be certain that all progenitors have emerged from the common envelope by the time the orbital period is this low. The value of $P_c$ can be determined by using
Equation~(\ref{eggleton_roche_lobe}) and the radius of the white dwarf
as determined by Lynden-Bell and O'Dwyer~\cite{lyndenbell01}. Adopting
the optimistic values of $N_{\rm X} = 10$ and $T_{\rm X} = 10^7 {\rm\ yr}$, and evaluating Equation~(\ref{detached_time}) gives $T_{\rm det} \sim 10^7 {\rm\ yr}$. Thus, we find $N_{\rm det} \sim 1 - 10$, which is within an order of magnitude of the numbers found through dynamical simulations (Section~\ref{subsubsection:monte-carlo_simulations}) and encounter rate estimations (Section~\ref{subsubsection:encounter_rates}).

Current production of ultracompact WD-NS binaries is more likely to arise through collisions of neutron stars with lower mass red giant stars near the current turn-off mass. The result of such a collision is a common envelope that will quickly eject the envelope of the red giant and leave behind the core in an eccentric orbit. The result of the eccentric orbit is to hasten the inspiral of the degenerate core into the neutron star due to gravitational radiation~\cite{peters63}. Consequently, $T_{\rm det}$ can be significantly shorter~\cite{ivanova05b}. Adopting a value of $T_{\rm det} \sim 10^6$ gives $N_{\rm det} < 100$.

Continuing in the spirit of small number statistics, we note that
there is one known radio pulsar in a globular cluster NS--NS binary
(B2127+11C) and about 50 known radio pulsars in the globular cluster
system as a whole (although this number may continue to
grow)~\cite{lorimer01}. We may estimate that NS--NS binaries make up
roughly $1/50$ of the total number of neutron stars in the globular
cluster system. A lower limit on the number of neutron stars comes
from estimates of the total number of active radio pulsars in
clusters, giving $N_{\rm NS^2} \sim 10^5$~\cite{kulkarni90}. Thus, we
can estimate the total number of NS--NS binaries to be $\sim 2000$. Not
all of these will be in compact orbits, but we can again estimate the
number of systems in compact orbits by assuming that the systems
gradually decay through gravitational radiation and thus
\begin{equation}
  \frac{N_{\rm compact}}{N_{\rm NS^2}} =
  \frac{T_{\rm compact}}{T_{\rm coalesce}}
  \label{NS_binary_ratio}
\end{equation}
where $N_{\rm compact}$ is the number of systems in compact orbits
($P_{\rm orb} < 2000 {\rm\ s}$), $T_{\rm compact}$ is the time spent
as a compact system, and $T_{\rm coalesce}$ is the typical time for a
globular cluster NS--NS binary to coalesce due to gravitational
radiation inspiral. Adopting the coalescence time of B2127+11C as
typical, $T_{\rm coalesce} = 2 \times 10^8 {\rm\ yr}$~\cite{prince91},
and integrating Equation~(\ref{detached_time}) for two $1.4 M_{\odot}$
neutron stars, we find $N_{\rm compact} \sim 25$. Again this value
compares favorably with the values found from encounter rate
estimations.

\newpage


\section{Prospects of Gravitational Radiation}
\label{section:gravitational_radiation}

A very exciting prospect for the observation of relativistic binaries
in globular clusters lies in the fact that they will be sources of
gravitational radiation. There is a phase in the evolution of most
relativistic binaries during which the orbital period is slowly
shrinking due to the emission of gravitational radiation. If the
binary is in a circularized orbit, the gravitational radiation will be
peaked strongly in the second harmonic of the orbital period, so
$f_{\rm gw} = 2f_{\rm orb}$. Gravitational radiation can be described
by the dimensionless strain amplitude $h_o$. Although the strength of
the gravitational radiation varies with the orientation of the binary,
an angle-averaged estimate of the signal strength is~\cite{lisa96}
\begin{equation}
  h_0 = 1.5 \times 10^{-21} \left(\frac{f_{\rm gw}}{10^{-3}
  {\rm\ Hz}}\right)^{2/3} \left(\frac{1 {\rm\ kpc}}{r}\right)
  \left(\frac{{\cal M}}{M_{\odot}}\right)^{5/3}.
\end{equation}
At a typical globular cluster distance of $r \sim 10 {\rm\ kpc}$ and
typical chirp mass of ${\cal M} \sim 0.5 M_{\odot}$, a relativistic
WD--WD or WD--NS binary with $P_{\rm orb} = 400 {\rm\ s}$ will have a
gravitational wave amplitude of $10^{-22}$. This value is within the
range of the proposed space-based gravitational wave observatory
LISA~\cite{lisa96}.

Many globular clusters lie off the plane of the galaxy and are
relatively isolated systems with known positions. The angular
resolution of LISA improves with signal strength. By focusing the
search for gravitational radiation using known positions of suspected
sources, it is possible to increase the signal-to-noise ratio for the
detected signal. Thus, the angular resolution of LISA for globular
cluster sources can be on the order of the angular size of the
globular cluster itself at $f_{\rm gw} > 1 {\rm\ mHz}$. Consequently,
the orbital period distribution of a globular cluster's population of
relativistic binaries can be determined through observations in
gravitational radiation. We will discuss the prospects for observing
each class of relativistic binaries covered in this review.

White dwarf--white dwarf binaries that are formed from a common
envelope phase will be briefly visible while the recently revealed hot
core of the secondary cools. These objects are most likely the
``non-flickerers'' of Cool {\it et al.}~\cite{cool98} and  Edmonds
{\it et al.}~\cite{edmonds99}. WD--WD binaries formed through exchange
interactions may very well harbor white dwarfs which are too cool to
be observed.  In either case, hardening through dynamical interactions
will become less likely as the orbit shrinks and the effective cross
section of the binary becomes too small. These objects will then be
effectively invisible in electromagnetic radiation until they are
brought into contact and RLOF can begin. During this invisible phase,
the orbital period is ground down through the emission of
gravitational radiation until the orbital period is a few hundred
seconds~\cite{benacquista99}. With a frequency of 1 to 10 mHz,
gravitational radiation from such a binary will be in the band of
LISA~\cite{lisa96}. There are $\sim 175$ such systems predicted from
encounter rates (see Table~\ref{encounter_populations}).

White dwarf--neutron star binaries that are expected to be progenitors
of the millisecond pulsars must pass through a phase of gravitational
radiation after the degenerate core of the donor star emerges from the
common envelope phase and before the spin-up phase begins with the
onset of mass transfer from the WD to the neutron star. The orbital
period at the onset of RLOF will be on the order of 1 to 2 minutes and
the gravitational wave signal will be received at LISA with a
signal-to-noise of $50-100$ at a frequency of around $20 {\rm\ mHz}$
for a globular cluster binary. Estimates of the number of such systems
range from $1 - 10$ for semi-empirical methods
(Section~\ref{subsubsection:semi-empirical_methods}) to $\sim 125$
from encounter rates (Table~\ref{encounter_populations}).

Binaries with significant eccentricity will have a spectrum of
harmonics of the orbital frequency, with the relative strength of the
$n$th harmonic for eccentricity $e$ given by~\cite{peters63}
{\arraycolsep 0.14 em
\begin{eqnarray}
  g(n, e) & = & \frac{n^4}{32}\bigg\{\left[J_{n-2}(ne) -
  J_{n-1}(ne) + \frac{2}{n}J_n(ne) + J_{n+1}(ne) -
  J_{n+2}(ne)\right]^2
  \nonumber \\
  & & \qquad + (1-e^2)\left[J_{n-2}(ne) \!-\! 2J_n(ne) \!+\!
  J_{n+2}(ne)\right]^2 \!+\! \frac{4}{3n^2}\left[J_n(ne)\right]^2
  \!\!\bigg\}, \nonumber \\
\end{eqnarray}}%
where $J_n$ is the Bessel function. The higher harmonics of
sufficiently eccentric binaries ($e > 0.7$) can be detected by LISA
even though the fundamental orbital frequency is well below its
sensitivity band of $1 - 100 {\rm\ mHz}$~\cite{benacquista02}.

Although the globular cluster population of NS--NS binaries is
expected to be quite small ($\sim 10$), they may have high
eccentricities. The binary pulsar B2127+11Cis an example of a NS--NS
binary in a globular cluster. In terms of the unknown angle of
inclination $i$, the companion mass to the pulsar is
$M_2\sin{i} \sim 1 M_{\odot}$ and its eccentricity is
$e = 0.68$~\cite{lorimer01}. These binaries may also be detectable by
LISA. If the globular cluster systems of other galaxies follow similar
evolution as the Milky Way population, these binaries may be potential
sources for LIGO as gravitational radiation grinds them down to
coalescence. With their high eccentricities and large chirp mass,
black hole binaries will also be good potential sources for
gravitational radiation from the galactic globular cluster
system~\cite{benacquista01b, benacquista02}.

The relatively close proximity of the galactic globular cluster system
and the separations between individual globular clusters allows for
the identification of gravitational radiation sources with their
individual host clusters. Although the expected angular resolution of
LISA is not small enough to allow for the identification of individual
sources, knowledge of the positions of the clusters will allow for
focused searches of the relativistic binary populations of the
majority of the galactic globular clusters. Armed with a knowledge of
the orbital periods of any detected binaries, concentrated searches in
electromagnetic radiation can be successful in identifying
relativistic binaries that may have otherwise been missed.

\newpage


\section{Summary}
\label{section:summary}

Relativistic binaries are tracers for the rich dynamical evolution of
globular clusters. The populations of these objects are the result of
an interplay between the gravitational dynamics of large $N$-body
systems, the dynamics of mass transfer, the details of stellar
evolution, and the effect of the gravitational field of the
galaxy. The gravitational dynamics of globular clusters can enhance
the population of short period binaries of main-sequence stars as well
as inject compact objects such as white dwarfs and neutron stars into
stellar binary systems. Once they are in such systems, the details of
stellar evolution and mass transfer in close binary systems govern the
likely end products of the dynamical interaction between the two
stars. Furthermore, most models of the evolution of the core of a
globular cluster rely on the gradual hardening and ejection of binary
systems to delay the onset of core collapse. The hardening of binaries
in the core of globular clusters will produce relativistic binaries,
but it will also eventually eject these systems as they gain larger
and larger recoil velocities in each subsequent encounter. The
threshold for ejection from a globular cluster depends both upon the
gravitational potential of the cluster itself and the gravitational
potential of its environment generated by the Milky Way. As the
globular cluster orbits the Milky Way, its local environment
changes. Consequently, if other dynamical processes (such as
gravothermal oscillations) do not dominate, the globular cluster's
population of relativistic binaries may also reflect the past orbital
history of the globular cluster.

Over the last decade, observational techniques and technology have
improved to the extent that significant discoveries are being made
regularly. At this point, the bottleneck in observations of binary
millisecond pulsars, low-mass X-ray binaries, and cataclysmic
variables is time, not technology. As these observational techniques
are brought to bear on more clusters, more discoveries are bound to be
made. In the next decade, the possibility of using gravitational wave
astronomy to detect relativistic binaries brings the exciting
possibility of identifying the populations of electromagnetically
invisible objects such as detached WD and NS binaries and black hole
binaries in globular clusters. These observations can only help to
improve the understanding of the complex and interesting  evolution of
these objects and their host globular clusters.

\newpage


\section{Acknowledgements}
\label{section:acknowledgements}

I would like to acknowledge the kind hospitality of the Aspen Center for Physics, and the anonymous referees whose extensive knowledge of the subject has helped fill in the gaps in my understanding. Extensive use was made of both the arXiv pre-print server and the ADS system. This work has been supported by NASA EPSCoR grant NCCW-0058, Montana EPSCoR grant NCC5-240, NASA Cooperative Agreement NCC5-579, and NASA APRA grant NNG04GD52G.

\newpage


\bibliography{newrefs}

\end{document}